\documentclass[11pt]{article}
\usepackage[margin=1.05in]{geometry}
\usepackage[T1]{fontenc}
\usepackage[utf8]{inputenc}
\usepackage{lmodern,microtype}
\usepackage{amsmath,amssymb,mathtools,amsthm}
\usepackage{booktabs,tabularx,array,enumitem,graphicx,caption,xcolor}
\usepackage[authoryear,round]{natbib}
\usepackage{hyperref,titlesec,placeins,pdflscape,flafter,afterpage,needspace}
\definecolor{accent}{RGB}{31,78,121}
\hypersetup{colorlinks=true,citecolor=accent,linkcolor=accent,urlcolor=accent,
 pdftitle={Adaptive Complementarity in Human-AI Systems: Architecture as a State-Shaping Choice},pdfauthor={Babak Heydari}}
\setlist[itemize]{leftmargin=1.3em,itemsep=2pt,topsep=3pt}
\setlist[enumerate]{leftmargin=1.45em,itemsep=2pt,topsep=3pt}
\titleformat{\section}{\large\bfseries}{\thesection}{.55em}{}
\titleformat{\subsection}{\normalsize\bfseries}{\thesubsection}{.5em}{}
\titlespacing*{\section}{0pt}{13pt}{5pt}
\titlespacing*{\subsection}{0pt}{10pt}{4pt}
\titlespacing*{\paragraph}{0pt}{8pt plus 1pt minus 1pt}{.75em}
\newcommand{\E}{\mathbb E}
\newcommand{\gov}{\mathsf g}
\newcommand{\G}{\mathcal G}
\newcommand{\doi}[1]{\href{https://doi.org/#1}{\nolinkurl{#1}}}
\newtheorem{proposition}{Proposition}
\newcolumntype{Y}{>{\raggedright\arraybackslash}X}

\newcommand{\HistoryRows}{%
AI-led & 62.18\% & \textbf{0.80000} & 0.78000\\
Independent work with combination & 69.45\% & 0.80000 & \textbf{0.85272}\\
}
\newcommand{\PolicyRows}{%
14 / Zero & 9.854322 & 9.854322 & 9.874582 & 0.020259\\
30 / Zero & 18.180627 & 18.524068 & 18.531942 & 0.007874\\
60 / Zero & 28.097874 & 29.332327 & 29.332679 & 0.000352\\
\addlinespace
14 / Common continuation & 40.000000 & 42.265488 & 42.265488 & 0.000000\\
30 / Common continuation & 40.000000 & 42.827269 & 42.827269 & 0.000000\\
60 / Common continuation & 40.000000 & 42.953679 & 42.953679 & 0.000000\\
}
\newcommand{\SelectiveRows}{%
0.5 & 0.5 & 0.79000 & AI-led & 0.82429 & Independent\\
0.6 & 0.4 & 0.79760 & AI-led & 0.83874 & Independent\\
0.8 & 0.2 & 0.81280 & Selective & 0.86766 & Selective\\
1.0 & 0.0 & 0.82800 & Selective & 0.89657 & Selective\\
}

\title{\vspace{-.8cm}\textbf{Adaptive Complementarity in Human--AI Systems:}\\[1mm]
\large Architecture as a State-Shaping Choice}
\author{Babak Heydari\\[-1mm]
\small College of Engineering and Network Science Institute, Northeastern University\\[-1mm]
\small MAGICS Lab}
\date{Preprint --- September 2026}
\begin{document}
\raggedbottom
\renewcommand{\topfraction}{.90}
\renewcommand{\bottomfraction}{.85}
\renewcommand{\textfraction}{.08}
\renewcommand{\floatpagefraction}{.85}
\maketitle
\vspace{-4mm}
\begin{abstract}
Human--AI interaction can improve current performance while changing the capabilities and relationships on which future performance depends. We develop adaptive complementarity, a framework for choosing interaction architecture with these state consequences in view. Access, information exposure, task allocation, timing, and communication can alter which arrangement will be valuable later; their settings can often be reset faster than the capabilities, search patterns, or conventions they create. Three mechanisms organize the argument: information exposure and collective search, delegation and capability evolution, and strategic interdependence and information governance. Their integration yields cross-mechanism implications, including conditions under which a loss of expertise heterogeneity increases the information differentiation required to preserve independent search. We distinguish strong human--AI complementarity from advantage over another workflow and from advantage over an evolving reference policy. A computational illustration examines scarce human review in a workflow whose success requires several specialized stages. Review develops human expertise and AI capabilities, changing where subsequent review is most valuable. Adaptive allocation improves net output over untailored procedures and the optimal predetermined calendar. An understandable priority rule derived from the adaptive solution retains essentially all of its gain: the procedure stays fixed while assignments respond to the capabilities that interaction creates. The framework directs evaluation toward the states present interaction creates, their consequences for later architectural fit, and the conditions under which observing and responding to them is worthwhile.

\end{abstract}

\begin{flushright}
\begin{minipage}{.78\textwidth}
\small\raggedleft
\textit{``Every experience is a moving force. Its value can be judged only on the ground of what it moves toward and into.''}\par
\smallskip
---John Dewey, \emph{Experience and Education} (1938), p.~38
\end{minipage}
\end{flushright}

\section{The complementarity puzzle}

Human--AI collaboration changes the conditions for its own success. Receiving AI suggestions, delegating tasks, and communicating with other participants can change what people learn, which ideas they explore, and how they rely on one another. Some of these changes strengthen future collaboration; others weaken it. This creates a design puzzle: an arrangement that improves performance today may change which arrangement will work best tomorrow. Designing human--AI systems therefore requires evaluating both what an interaction accomplishes and what it leaves behind.

A field experiment in education makes this puzzle concrete. Students using a conventional GPT-4 interface performed better while assistance was available, but subsequently performed worse without it than students in a control group. A tutoring interface using the same underlying model, with teacher-informed guidance, largely mitigated this learning penalty \citep{bastani2025guardrails}. How assistance was provided therefore affected both the work students could complete with AI and the capabilities they carried into later tasks. An evaluation limited to assisted performance would miss the second outcome.

Related tensions appear in creativity, decision-making, and coordination. When writers used AI suggestions, their individual stories improved, but the stories became more alike: better individual output came with less variety across the group \citep{DoshiHauser2024}. Designers shown AI-generated examples were more likely to fixate on those examples, limiting their exploration of alternatives \citep{WadinambiarachchiEtAl2024}. In decision-making, adding explanations to AI recommendations increased acceptance without improving people's ability to distinguish correct recommendations from incorrect ones. Interfaces that required more deliberate engagement reduced overreliance, but demanded additional effort \citep{BansalEtAl2021Explain,BucincaEtAl2021}. Communication among AI agents presents a related tension: it can help them coordinate, but coordination that benefits the participants need not benefit others affected by their decisions \citep{dafoe2020cooperative}.

We use \emph{interaction architecture} to describe how assistance is organized: who has access to AI, what information they see and when, how tasks and decision rights are allocated, and how participants communicate. These are choices that a designer can often adjust more quickly than the capabilities and relationships they shape. Removing an AI suggestion from an interface, for example, need not remove the ideas it has already anchored in a user's thinking. Reassigning a task to a person does not immediately restore the skill needed to perform it.

We develop \emph{adaptive complementarity} as a framework for choosing interaction architecture with these consequences in view. Three mechanisms organize the analysis: how information exposure shapes collective search, how delegation shapes capabilities, and how information governance shapes cooperation and competition when participants' decisions affect one another. Their connections matter as much as their separate effects. Suppose, for example, that repeated delegation makes team members' expertise more similar. If differences in expertise previously helped them explore independently, preserving the same degree of independent search may now require giving them more differentiated information. A choice about who performs tasks can therefore change what is required of a later choice about who sees which information.

This perspective also requires clarity about what counts as successful collaboration. \emph{Augmentation} means that people perform better with AI assistance than without it. \emph{Strong complementarity} sets a higher performance threshold: the combination must outperform both the human alone and the AI alone under comparable conditions. A meta-analysis of 106 experiments found that human--AI combinations substantially outperformed humans alone but, on average, underperformed the better standalone participant, with considerable variation across tasks \citep{Vaccaro2024}. For binary prediction tasks, formal analysis shows that even when predictions are calibrated, deterministic rules for combining them cannot guarantee that the combination outperforms both participants across all admissible settings \citep{PengGargKleinberg2025}. Differences in capabilities therefore create opportunities whose value depends on how those capabilities are combined.

Strong complementarity is nevertheless only one relevant comparison. An institution may need to know whether a new workflow improves on an existing one, or whether a policy for organizing interaction produces better outcomes over time. These questions require different benchmarks. The distinction becomes especially consequential when interaction changes capabilities: an apparent improvement relative to a weakening benchmark can conceal declining performance in the assisted system itself. Comparisons over time must also allow each alternative to produce its own learning history. Testing what a person can do after using AI, for example, does not establish what that person would have learned without it.

Adaptive architecture need not require continual redesign of the procedure that governs it. The framework separates three claims. First, an earlier arrangement may change a condition that matters for future performance, such as expertise or reliance; we call this \emph{state creation}. Second, that change may alter which later arrangement works best; this is \emph{history-dependent fit}. Third, measuring the evolving condition and responding to it may improve outcomes enough to justify the costs; this is the \emph{operational value of feedback}. Each claim requires its own evidence. An organization can discover how allocation should respond to developing capabilities and embody that response in a stable priority rule. The rule remains unchanged, while the assignments it produces adapt to experience. Determining the value of this response and finding a practical way to implement it are complementary design tasks.

\paragraph{Positioning.}
Human--AI teaming research provides the immediate foundation. Work on teaming examines how complementary cognitive functions, shared knowledge, roles, and training support joint performance \citep{GonzalezEtAl2026}. Research on bidirectional alignment and human--AI coevolution considers how people and AI change in response to one another, including effects beyond a single pair \citep{ShenEtAl2025,PedreschiEtAl2025}. Advising and routing approaches make assistance conditional on the user or task \citep{NotiChen2023,AminEtAl2026}. Delegation models show how early choices can shape later capabilities by changing opportunities for practice, even when participants pursue a common performance objective \citep{huang2026geometry}. We connect these perspectives through an explicit account of the architectural choices being made, the conditions they create, and how those conditions affect subsequent choices.

\paragraph{Contribution.}
The paper makes three organizing moves. First, it distinguishes what a designer can change directly from the capabilities, task conditions, and relationships that develop through interaction, while recognizing practical limits on control and observation. Second, it connects information exposure, delegation, and strategic information governance, identifying how consequences in one area can change design requirements in another. Third, it separates the comparisons needed for evaluating strong complementarity, advantage over another workflow, and advantage over time, together with empirical designs that separate the effects of past interaction from the value of responding to them. The contribution lies in integrating these ideas into a framework for human--AI design and evaluation, with formal analysis clarifying the conditions under which particular implications follow. The computational illustration connects this framework to implementation: discovering an adaptive allocation policy can reveal meaningful priorities that an organization can apply through a simple operating rule.

Section~\ref{sec:framework} develops the framework and its evaluation criteria. Sections~\ref{sec:mechanisms} and~\ref{sec:synthesis} examine the three mechanisms and their connections. Section~\ref{sec:demo} examines review across interdependent workflow stages, identifies the value of adapting to realized learning, and derives an understandable rule from the adaptive solution. Section~\ref{sec:empirical} develops their empirical evaluation and explains how policy discovery can inform simpler organizational rules. The appendices provide the governance formulation, derivations, the workflow model specification, a comparison with independent outputs, and a separate illustration of how learning histories change architectural fit at equal aggregate human competence.

\section{Architecture, state, and evaluation}
\label{sec:framework}

\subsection{What architecture changes}

Capability differences create the opportunity for collaboration; architecture shapes how those differences enter the work. Repeated use then changes the conditions under which the next decision is made. Describing that process requires separating two things that are easy to conflate: what a designer can change now, and what the system has become through prior interaction. Three components describe the latter:
\begin{equation}
X_t=(A_t,P_t,Z_t).
\label{eq:state}
\end{equation}
\textbf{Agent state} $A_t$ includes human and AI capabilities, learned representations, and relevant behavioral tendencies.

\textbf{Problem state} $P_t$ includes epistemic structure---the conditions for finding and evaluating solutions---and strategic structure---the incentives and consequences of participants' interdependent choices. Epistemic features include uncertainty, ruggedness (multiple local peaks that can trap search), decomposability (how far a task can be divided into relatively independent parts), and remaining search opportunities. Strategic features include coordination demands, competitive conditions, and externalities: effects on others not fully reflected in a participant's own payoff.

\textbf{Relational state} $Z_t$ records conditions produced by interaction, such as reliance, correlated search, shared routines, or strategic conventions. Epistemic and strategic structure can interact: common information may improve individual decisions while imposing a collective cost by reducing independent exploration.

Interaction architecture $I_t$ specifies access, information exposure, task allocation, decision rights, timing, communication, and connectivity. The distinction between architecture and state is operational, and the clearest way to see it is to try to undo something. A team can change whether members see a common AI suggestion before producing their own ideas; it cannot directly reset the shared anchors those suggestions have already created. Restricting delegation withdraws an affordance; it does not immediately restore a skill. Realized delegation and verification effort are behaviors induced by architecture and state, rather than quantities that a designer can necessarily set.

Architecture is often only partly governable: a designer can control some features of interaction without determining all channels or behavioral responses. An organization may control its official interface while users retain side channels; a regulator may constrain communication without choosing an agent's strategy. In the compact formalism below, feasible architectural choices incorporate these constraints, while performance and state transitions account for the behavior those choices induce. Appendix~\ref{app:governance} makes the governed subset, response selection, and partial observability explicit.

A fixed rule need not produce a fixed architecture. A constant setting $I_t=\bar I$ holds the governed choice unchanged. Fixed allocation probabilities randomize choices without responding to time or state. A predetermined schedule $I_t=\iota_t$ changes choices with time, whereas a state-responsive rule $I_t=\pi(X_t,t)$ conditions them on observed circumstances. The function $\pi$ can be specified before operation and remain unchanged. Adaptive complementarity concerns the response of architectural choices to consequential state, rather than requiring continual redesign or retraining of the rule that produces them. The value of that response and the complexity of implementing it are separate questions: a simple rule can preserve a substantial adaptive benefit. When state is partly observed, the rule uses available history or a belief instead of $X_t$.

The dynamic premise is
\begin{equation}
X_{t+1}=F(X_t,I_t,\varepsilon_{t+1}).
\label{eq:transition}
\end{equation}
This transition can improve or degrade different state components. Delegation can reduce execution practice while allowing supervisory competence to develop. Differentiated information can preserve independent representations but slow the spread of a useful discovery. Communication can build beneficial conventions or harmful coordination. Problem state can also change endogenously: a discovery may expose a new dependency that requires joint investigation, while recommendations can shift competitive positions and thereby alter the returns to later recommendations. The framework evaluates which changes matter for the task and horizon at hand. Its state description need not consist of independent coordinates; search overlap, for example, may be derived from agents' learned representations. A particular model should retain only the components its mechanism requires.

\subsection{Three comparisons that answer different questions}

Let $w^{\G}(X,I)$ denote system value under a specified stakeholder objective $\G$: what counts as a better outcome and whose interests enter the assessment. It may aggregate accuracy, effort, reliability, or welfare, provided the tradeoffs are stated. Participating agents can pursue private objectives $u_i$ that differ from $w^{\G}$. Coordinating agents might, for example, increase their principals' profits while making consumers worse off. A rise in profits, agreement, or cooperation therefore need not increase the evaluator's value. We suppress $\G$ after fixing this evaluation boundary; when scalar aggregation is inappropriate, constraints or multiple objectives can replace it.

The complementarity puzzle motivates architectural design, but three comparisons must remain distinct (Table~\ref{tab:benchmarks}). For a human--AI dyad, contemporaneous strong complementarity is
\begin{equation}
C_t=w(X_t,I_t)-\max\{w_H(X_t),w_{AI}(X_t)\}.
\label{eq:complementarity}
\end{equation}
The constituents are evaluated under the same objective and comparable task and resource conditions. A broader workflow comparison instead measures \emph{architectural advantage}, $M_t=w(X_t,I_t)-b(X_t;\mathcal B)$, over a prespecified reference class $\mathcal B$. We reserve $C$ for constituent-based complementarity and use $M$ for the more general gap. Improvement over a status quo alone does not establish strong complementarity.

\begin{table}[htbp]
\small\centering
\begin{tabularx}{\textwidth}{@{}p{.23\textwidth}YY@{}}
\toprule
\textbf{Comparison} & \textbf{Reference} & \textbf{Interpretation}\\
\midrule
Strong complementarity & Better standalone constituent at the current state & Does joint performance exceed either constituent's independent performance?\\
\addlinespace
Architectural advantage & A specified alternative workflow evaluated at the current state & Does this arrangement improve on that feasible alternative?\\
\addlinespace
Trajectory policy advantage & A prespecified policy class, each policy generating its own state trajectory & Does this interaction policy create more total value from a common initial condition?\\
\bottomrule
\end{tabularx}
\caption{The evaluation objective, resource conditions, and reference class are part of the comparison. Only the first row necessarily tests the opening strong-complementarity criterion.}
\label{tab:benchmarks}
\end{table}

The MASAI mammography trial makes the architectural comparison concrete. Its AI-supported workflow used AI to help determine which examinations received one or two radiologist readings and to support detection; the control workflow used two radiologists without AI. The initial safety analysis reported a 44.3\% reduction in screen-reading workload with a similar cancer detection rate \citep{LangEtAl2023MASAI}. Follow-up found higher screening sensitivity with unchanged specificity and met the prespecified non-inferiority criterion for interval cancers---cancers diagnosed between screening rounds \citep{GommersEtAl2026MASAI}. These findings support an institutional workflow comparison. Establishing strong complementarity would additionally require comparable evaluations of the standalone constituents. How sustained use of either workflow changes radiologists' capabilities remains a separate longitudinal question.

There is a measurement trap here: a system can appear to improve relative to its benchmark simply because the benchmark is deteriorating faster. For either gap $M_t=w_t-b_t$ or its constituent-based special case $C_t$, the accounting identity is
\begin{equation}
M_{t+1}-M_t=(w_{t+1}-w_t)-(b_{t+1}-b_t).
\label{eq:benchmark}
\end{equation}
The gap rises while system value falls if the benchmark deteriorates faster. This is a measurement implication, not an independent behavioral theorem. For instance, assisted performance falling from 90 to 85 alongside unaided human performance falling from 80 to 70 increases the human-reference gap from 10 to 15. With AI-alone performance fixed at 75, the strong-complementarity gap instead stays at 10: the AI creates a floor under the better-constituent benchmark. The distinction prevents a human-reference deterioration from being misreported as an increase in strong complementarity.

Taking the AI away and measuring what someone can do reveals their current capability after an assisted learning history. It does not reveal what they would have been able to do had they never used it, because that alternative history could have involved different practice. Comparisons across policies must therefore let each alternative produce its own capabilities and relational state. Evaluating every alternative on the focal policy's state would omit exactly the part of the difference that interaction history created.

\afterpage{\clearpage
\begin{landscape}
\begin{figure}[p]
\centering
\includegraphics[width=.92\linewidth]{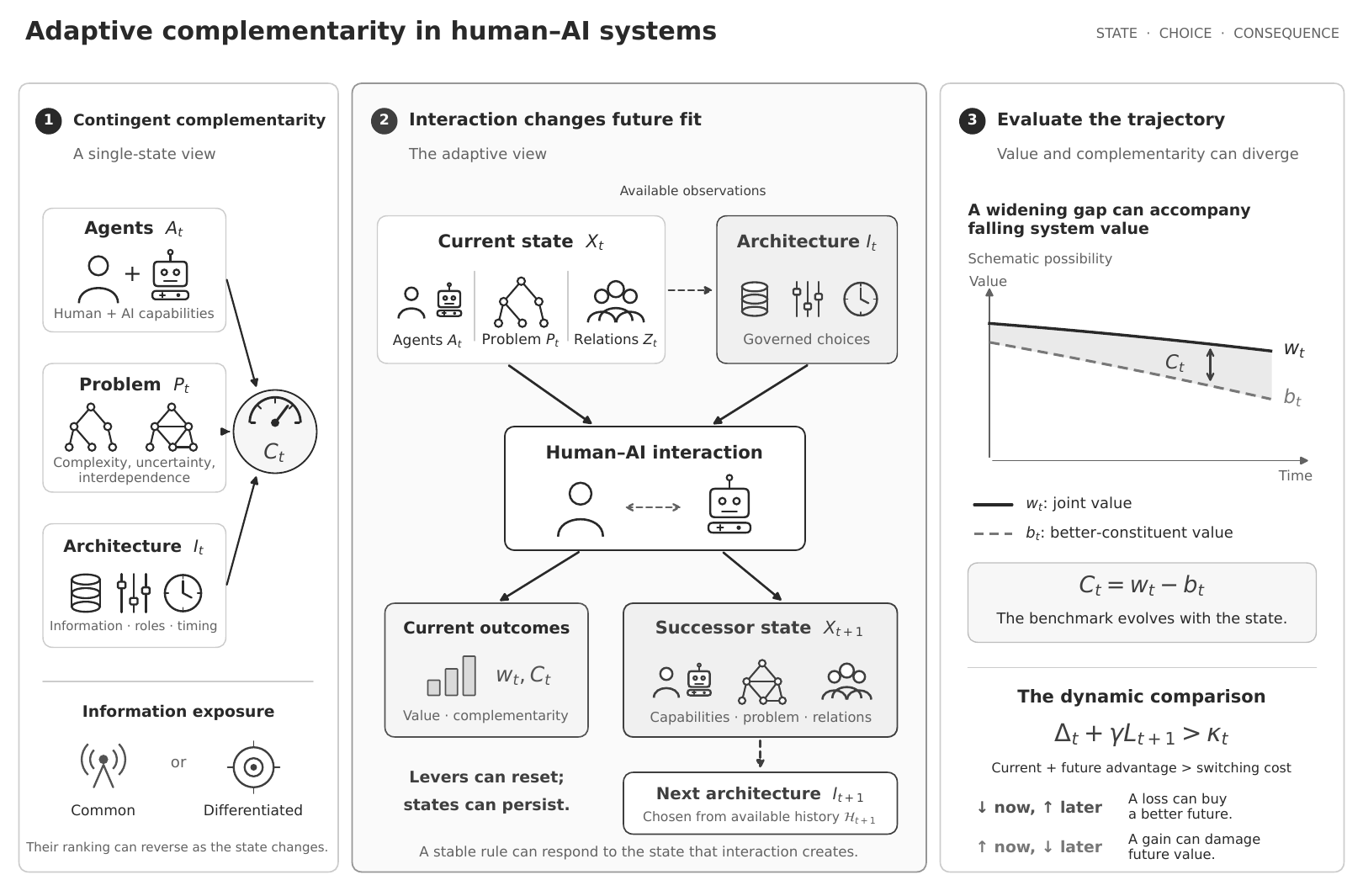}
\caption{\textbf{Adaptive complementarity.} (1) Architectural fit depends on agents and problem structure. (2) Interaction creates outcomes and a successor state. A stable rule can choose the next architecture using available history $\mathcal H_{t+1}$. (3) The better-constituent gap $C_t=w_t-b_t$ can widen while joint value falls. The schematic shows a range in which the human reference remains dominant; a fixed AI reference would eventually bound the benchmark's decline. The choice criterion compares a candidate with the incumbent, including current switching cost $\kappa_t$ (Eq.~\eqref{eq:criterion}). These are conceptual trajectories, not computational results.}
\label{fig:framework}
\end{figure}
\end{landscape}}

\subsection{A compact dynamic comparison}
\label{sec:formal}

Fix the stakeholder objective and behavioral-response model. Let $\pi$ be a policy: a rule for choosing feasible architecture from the information available. The discount factor $\gamma\in(0,1)$ weights future performance, and $\kappa(I,I^-)$ denotes the direct cost of moving from the previously implemented architecture $I^-$ to $I$, with $\kappa(I,I)=0$. The objective sums expected performance over time, subtracting the direct costs of changing architecture:
\begin{equation}
J_\kappa(\pi\mid x_0,i_{-1})=
\E_\pi\!\left[\sum_{t=0}^{\infty}\gamma^t
\{w(X_t,I_t)-\kappa(I_t,I_{t-1})\}\right].
\label{eq:objective}
\end{equation}
Including switching costs makes the inherited architecture part of the state. Let $V^*(x,i^-)$ be optimal continuation value: the best expected discounted performance, net of switching costs, available from state $x$ and inherited architecture $i^-$. For two current choices $I^1$ and $I^0$, define $\Delta_t=w(X_t,I^1)-w(X_t,I^0)$ and
\begin{equation}
L_{t+1}^{1,0}=\E\!\left[V^*(X_{t+1}^1,I^1)-V^*(X_{t+1}^0,I^0)\mid X_t\right].
\label{eq:continuation}
\end{equation}
$\Delta_t$ measures the difference in immediate performance; $L_{t+1}^{1,0}$ measures the difference in future value created by the two choices, allowing later choices to be optimized. The superscripts name the choices in the order compared. Reversing that order reverses the sign. The exact full-information comparison is
\begin{equation}
\boxed{I^1\succ I^0\quad\Longleftrightarrow\quad
\Delta_t+\gamma L_{t+1}^{1,0}>
\kappa(I^1,I_{t-1})-\kappa(I^0,I_{t-1}).}
\label{eq:criterion}
\end{equation}
If $I^0$ is the incumbent, the right side is simply the cost of switching to $I^1$. Future switching costs are already included in $V^*$. Persistent capability or relational consequences enter $L$, avoiding double counting them as direct adjustment costs.

Equation~\eqref{eq:criterion} applies standard dynamic-programming logic \citep{Puterman1994} to human--AI architectural choice: weigh what an arrangement pays now against the value of the state it leaves behind, net of the cost of changing. Read in one direction, an immediate loss can be justified by learning or coordination worth enough to offset it. Read in the other, an immediate gain can be rejected when what it creates is costly enough. The equation disciplines that comparison; it does not supply a new optimization method.

With zero switching costs, a discrete pair's \emph{switching locus} is the set of states where $\Delta+\gamma L=0$: the alternatives have equal current-plus-continuation value. A reversal requires reachable states on both sides. An immediate-payoff crossing, $\Delta=0$, need not be a dynamic action-value crossing; the same investment may remain optimal because its continuation benefits are sufficiently large. Neither state change nor the notation alone guarantees a reversal. Switching costs can create inaction regions in which retaining the incumbent is preferable; continuously adjustable architecture can change smoothly without a discrete switch.

The full-information expression characterizes the tradeoff when the relevant state is known. An implementable policy generally uses observation history or a belief over latent state, such as skill or reliance that cannot be observed directly; Appendix~\ref{app:governance} gives the corresponding formulation. Noisy measurements, limited control, and costly adjustment can make a stable architecture preferable even when a fully informed policy would change it.

For a prespecified benchmark policy class $\Pi^B$, define trajectory policy advantage as
\begin{equation}
\mathcal A_{\rm traj}(\pi)=J_\kappa(\pi\mid x_0,i_{-1})-
\sup_{\pi^B\in\Pi^B}J_\kappa(\pi^B\mid x_0,i_{-1}).
\label{eq:trajectory}
\end{equation}
Both sides start from the same initial conditions and generate their own subsequent states. Initial setup differences, if relevant, must be charged consistently. When $\Pi^B$ represents meaningful human-only and AI-only policy alternatives, this supports a trajectory-level complementarity comparison; otherwise it is a policy advantage. The distinction is retained in the computational illustration.

\section{Three mechanisms}
\label{sec:mechanisms}

Each mechanism traces the same loop. An architectural choice changes how participants behave; that behavior leaves something behind---overlapping search, a portfolio of skills, a pattern of strategic expectation---and what it leaves behind can change which architecture is worth choosing next. The three differ in the behavioral process that drives them, and therefore in the state that accumulates and how readily it can be undone. Information exposure shapes collective search. Delegation shapes capability and reliance. Strategic information governance shapes cooperation and competition. The three can operate at once and do not exhaust the possibilities. Each subsection below identifies the behavioral process, the state it creates, the predictions that follow, and the measurements that separate an immediate effect from a consequence carried forward.

\subsection{Information exposure and collective search}
\label{sec:information}

\paragraph{Problem and evidence.}
The same suggestion that improves one person's next decision can narrow what a group collectively explores. Common AI output helps people pursue a promising candidate; it also points them toward the same candidate. When the answer that matters lies outside a plausible but incomplete consensus, the cost of that convergence is shared across the group and may not be fully reflected in any individual's decision.

The evidence separates these two levels. In the story-writing experiment of \citet{DoshiHauser2024}, AI-generated ideas improved individual story evaluations while making the stories more similar to one another: better individual output, less variety across the group. In AI-supported visual ideation, exposure to generated examples can increase fixation, limiting exploration of alternatives \citep{WadinambiarachchiEtAl2024}. Exposure is also configurable rather than given: differentiated AI search and selective idea sharing offer ways to combine assistance with independent work \citep{boussioux2024crowdless,GraftMind2026}. What these studies establish is a contemporaneous effect on output. Whether the effect persists after the AI is withdrawn, and whether preserving independence pays over a longer search, remain questions for longitudinal designs.

The underlying tradeoff predates AI: efficient communication spreads good solutions rapidly and can produce premature convergence on complex landscapes \citep{lazer2007network}. AI provides a channel for common exposure \emph{without human-to-human connectivity}. People who never speak to one another may still draw on overlapping suggestions, retrieval contexts, or representations. Network topology and informational commonality can therefore vary independently, and a sparse communication network is no guarantee of independent search when everyone consults the same source.

\paragraph{Architecture and state transition.}
Let $I^C$ provide greater common exposure and $I^D$ more differentiated exposure. The governed levers include prompting, retrieval context, private-first ideation, delayed pooling, and selective routing. Let $\rho_t\in Z_t$ measure overlap in search trajectories, with larger values indicating less independence. Its evolution depends on past overlap, realized exposure, agent heterogeneity, and the problem. When earlier exposure creates persistent anchors or shared representations, changing what people are shown need not immediately restore independent search: the shared frame can outlive the material that produced it.

Suppose common exposure carries a current exploitation advantage $g_t\geq0$ and raises expected next-state correlation by $\Delta\rho_{t+1}>0$ relative to differentiated exposure. If continuation value is locally decreasing in correlation, a first-order comparison can be written
\begin{equation}
Q_t(C)-Q_t(D)=g_t-\nu_{t+1}\Delta\rho_{t+1}+R_t,
\label{eq:information}
\end{equation}
where $Q_t$ includes current and continuation value, $\nu_{t+1}\geq0$ is the discounted local marginal value of search independence, and $R_t$ collects higher-order terms and other state effects. The expression weighs an immediate benefit against the future value of the independence it consumes. Ignoring $R_t$, differentiation is preferred when $\nu\Delta\rho>g$; if $|R_t|\leq\epsilon$, the stronger inequality $\nu\Delta\rho-g>\epsilon$ suffices. The quantity $\nu$ must be measured or modeled independently. It cannot be inferred from the observation that differentiation happened to work.

\paragraph{Design implications and boundary conditions.}
Rankings can reverse when changes in problem structure make the future value of independence outweigh the current benefit of common exposure, or vice versa. The local threshold predicts a reversal only when the remainder is small enough to preserve the sign of the comparison. Ruggedness, remaining search opportunity, and the risk that a shared premise fails are candidate determinants, but they are not interchangeable proxies: complexity need not always raise the value of diversity. Controlled tasks can vary the payoff to an additional independent starting point, the presence of local traps, or the conditions under which a previously useful solution becomes obsolete. Related computational and behavioral work links the value of independent search to landscape structure and documents divergence between private reliance and collective exploration \citep{HeydariEtAl2026SMJ,PadheeHeydari2024}.

As search resolves uncertainty, the value of further independence may fall and the preferred architecture may shift toward pooling---provided that offsetting movements in $g_t$ or in the induced correlation gap are not large enough to reverse the effect. New evidence or a regime change can raise the value of independence again. The direction of adjustment therefore tracks current search conditions rather than following a universal ``explore early, exploit late'' schedule.

The mechanism can also generate an externality. A user may receive an immediate benefit from common information while sharing the cost of reduced independent search with others. A net collective loss occurs when the loss imposed on others, after crediting any benefits from the user's improved work, exceeds the user's own gain. That loss can include forgone search option value---the value of retaining alternative paths to a solution. Reduced independence alone establishes neither a net collective loss nor a divergence in incentives. The externality arises when participants' decisions do not fully account for their effects on others' search opportunities.

\paragraph{Diagnostic and intervention.}
Low diversity is not itself a diagnosis. Independent searchers may converge because the evidence supports the same answer. Useful measurement combines trajectory similarity, discovery of previously unexplored regions, shared evidence, and unresolved search opportunity; a decline in novelty while consequential uncertainty remains is more informative than similarity alone. Comparisons should also separate similarity in the AI inputs people receive from convergence in how they interpret demonstrably different inputs.

Where valuable independence is being lost, interventions can differentiate retrieval, assign distinct search roles, delay common suggestions, or share group information selectively; pooling becomes appropriate when spreading useful discoveries matters more. A longitudinal test randomizes early exposure, later imposes the same architecture on everyone, and asks whether prior common exposure delays recovery once the task rewards renewed independent search. A difference that survives equalized information provision is evidence that the earlier architecture created a consequential state.

\subsection{Delegation, capability, and reliance}
\label{sec:delegation}

\paragraph{Problem and evidence.}
Delegation allocates two things at once: current work, and opportunities to learn. Assigning a task to AI shapes which capabilities a person exercises, which feedback they receive, and what they become able to contribute later. Those changes can then alter the relative value of human involvement and further delegation. The classic automation concern is one instance: people remain responsible for exceptional conditions while losing the practice needed to handle them \citep{Bainbridge1983,endsley1995outofloop}.

Generative AI extends this from execution to generation, evaluation, and problem framing. The tutoring comparison in the introduction shows what is at stake: the same underlying model, delivered through different interaction architectures, supported different subsequent human performance \citep{bastani2025guardrails}. Adaptive advising and advice timing likewise show that exposure changes engagement, not just the information available for a decision \citep{NotiChen2023,YinEtAl2025Timing}.

\paragraph{Architecture and state transition.}
Human capability is a vector---execution, oversight, exploration, framing---whose components need not move together. Delegation may reduce execution practice while helping a person develop supervisory expertise. AI capability can also evolve through human feedback, and reciprocal learning can be deliberately structured \citep{teeni2023reciprocal}. We distinguish capability $h_t$, emergent reliance $r_t$, and realized delegation $d_t$: what a person can do, their pattern of dependence on AI, and how much work they actually hand over. A governed rule influences $d_t$ without directly choosing the other two, and as capability and reliance evolve the same rule can induce different behavior.

Composition matters alongside level. What a person contributes to joint performance depends both on how much they know and on how their knowledge complements the AI's capabilities. Two people with identical unaided accuracy can therefore offer different potential contributions to the same collaboration. Work on training AI for team performance already distinguishes standalone accuracy from usefulness to a particular partner \citep{WilderEtAl2020,BansalEtAl2021Team}; the dynamic question is how architecture shapes that usefulness by determining which problems receive independent human effort and corrective feedback. Appendix~\ref{app:computation} isolates this composition effect: two learning histories produce the same $68\%$ unaided human accuracy but different coverage of the AI's failures, reversing the preferred next workflow. Section~\ref{sec:demo} examines a further consequence of capability development. When successful work requires several specialized stages, scarce review develops human expertise and improves the corresponding AI. Realized learning changes which stage most benefits from the next review, while the value of an improvement depends on the other stages' capabilities.

Two further feedbacks concern declining capability or engagement, and they are easily confused. In a \emph{practice--capability feedback}, delegating reduces opportunities to exercise a skill; where that skill depreciates through non-use, independent execution becomes less effective or more costly, which makes further delegation more attractive. Each decision can be justified by current comparative advantage while jointly deepening the capability gap on which later decisions rest. \citet{huang2026geometry} formalize coupled delegation and skill dynamics that produce persistent reliance and adverse long-run outcomes under specified assumptions.

In an \emph{incentive--engagement feedback}, what changes first is the return to exercising a capability rather than the capability itself. As AI becomes more reliable, verification catches fewer errors; holding checking costs and error consequences fixed, its immediate expected return declines. People may verify less while remaining fully able to verify. Should reduced verification later erode that ability, the two feedbacks couple. Expertise-atrophy and endogenous-oversight models formalize closely related channels \citep{siderius2026lose,GuLiZhu2026}.

The distinction guides the intervention. Where capability is intact and effort has fallen, increasing the reward or responsibility for verification may restore effort. Capability lost through non-use may instead require practice or retraining. A change in the delegation default can help by restoring the relevant engagement or learning opportunities.

\paragraph{Design implications and boundary conditions.}
Let $I^D$ favor greater delegation and $I^H$ retain more direct human involvement. The dynamic comparison asks whether delegation's immediate efficiency gain compensates for the difference in future capabilities the two arrangements produce. Retaining involvement can be worthwhile when it develops or preserves expertise later collaboration requires; greater delegation can be worthwhile when it frees people to acquire more valuable expertise or supply better feedback to the AI. Equation~\eqref{eq:criterion} evaluates both through the continuation value of the resulting state.

Preservation becomes more valuable when a capability is expensive to reacquire, hard to substitute, or disproportionately useful in states such as AI failure or environmental change. These comparative statics hold with the current cost of maintenance and other effects controlled; greater uncertainty alone is not sufficient. Retained capability has option value when it enables a useful response to future contingencies; its benefits can be much larger in some states than others. Oversight and problem-framing capabilities can matter out of proportion to their frequency of use, because they influence which subsequent actions are considered or challenged.

Human involvement can help create the conditions for greater delegation later. Performing tasks and reviewing AI outputs may develop supervisory expertise, improve the AI through corrective feedback, or reveal which tasks each participant handles best. Once these gains have developed, some work previously requiring direct human involvement may be delegated more effectively. Extensive delegation from the outset could prevent the learning that makes this later arrangement successful. Continued involvement remains valuable where practice is needed to sustain those capabilities. The design question is therefore how to allocate work both to develop the capabilities needed for future collaboration and to use them effectively as they evolve.

\paragraph{Diagnostic and intervention.}
Withdrawal performance, seeded-error detection, relearning curves, and verification behavior measure different aspects of the mechanism. Withdrawal tests assess unaided performance after exposure. Seeded errors---deliberately inserted mistakes---test oversight, though detection also depends on effort. Relearning curves measure how quickly lost performance returns and at what cost. Verification behavior, assessed under comparable correctness and task difficulty, helps characterize reliance.

Falling verification effort alongside stable detection ability is \emph{consistent with} an incentive--engagement pathway but does not identify it: fatigue, task mix, and perceived responsibility can produce similar patterns. A stronger design randomizes verification incentives while holding task conditions fixed and tests ability under standardized, motivated conditions, with repeated assessment to separate an immediate effort response from a later capability change. Independent human-first trials, selective routing, structured hints, and responsibility rotation can then be evaluated against the particular capability or pattern of reliance each is meant to preserve or change.

\subsection{Strategic interdependence and information governance}
\label{sec:strategic}

\paragraph{Problem and evidence.}
When participants' outcomes depend on one another's actions, what each knows about the others is not merely informative---it is strategic. Disclosing a rival's behavior, a partner's history, or a customer's response can change how participants act. Their actions can create reputations, conventions, and competitive positions that change what the next disclosure accomplishes. Institutions configuring AI agents on behalf of human principals---the people or organizations they represent---must therefore weigh both the cooperation or competition they induce now and the conditions they leave for later.

Strategic information governance connects research on cooperative AI with the economics of information design. Cooperative AI examines how artificial agents can support cooperation among human and machine participants \citep{dafoe2020cooperative}. Information design treats what participants learn as a means of influencing their choices without directly selecting their actions \citep{BergemannMorris2019Information}. In multi-agent reinforcement learning, this problem is explicitly sequential: information affects subsequent trajectories, and recipients need reasons to act on it \citep{LinEtAl2023Information}. Experiments with LLM populations also show that interaction can generate shared conventions and collective biases absent in isolated agents \citep{AsheryEtAl2025Conventions}. Together, these literatures motivate examining both how information governs present behavior and how interaction changes the conditions for subsequent governance.

A computational study of \emph{adaptive information modulation} (AIM) makes the loop visible \citep{ChenEtAl2026AIM}. LLM agents play repeated Prisoner's Dilemma games on a fixed interaction network, where each agent has an individual incentive to defect in any isolated round although mutual cooperation pays more than mutual defection. A governor, trained by reinforcement learning using rewards for its interventions, selects which historical information each agent receives; payoffs are unchanged and agents choose their own actions. The information network---who observes what---is thus governed separately from who interacts with whom, and the learned policy improves cooperation and aggregate payoffs relative to the fixed-disclosure baselines tested.

The learned policy is instructive about why disclosure's value depends on the history being disclosed. In a poorly cooperating neighborhood, revealing the cooperation rate can reinforce defection. Early on the governor sometimes supplies only the latest pairwise actions, including different information for the two participants; neighborhood information becomes dominant as cooperation develops. This pattern is consistent with an architectural feedback: disclosure shapes behavior, behavior becomes the historical record, and that record changes what the next disclosure conveys. The governor is trained offline and then applies a state-conditioned policy, so adaptive governance need not mean continual retraining.

A related network-intervention model allows a governor to change one interaction link per step while agents imitate more successful neighbors; faster imitation makes cooperation harder to preserve in the reported Prisoner's Dilemma settings \citep{ChenHeydari2025Network}. The distinction matters wherever an institution can govern disclosure more readily than it can govern who interacts with whom. Strategy imitation here is distinct from the capability acquisition considered in Section~\ref{sec:delegation}.

Recommendation can shape the strategic environment without any communication between rivals. In a platform-competition model, \citet{CaoEtAl2026Strategic} treat recommendation emphasis across customer segments as a sequential design choice: recommendations move customer choices and market shares, and those shares then condition later customer responses and the returns to future recommendations. Each decision therefore alters both current performance and the position from which the next is made, and policy effectiveness depends on that inherited position as well as on the opponent's behavior. The analysis studies one learning platform against fixed-policy opponents; it does not establish simultaneous learning by competitors or evolving individual preferences.\footnote{We refer to the revised manuscript under review. Its earlier public version is \citet{CaoEtAl2025Attention}; the revision's robustness results discussed in Section~\ref{sec:empirical} should not be attributed to that earlier version.}

Greater coordination does not by itself establish greater value under the evaluator's objective. A minimal channel increased cooperation among LLM agents in a four-player Stag Hunt, a game in which cooperating is rewarding when others cooperate too \citep{MadmounLahlou2026}, and a working-paper experiment reports more coordinated pricing in a competitive market simulation \citep{giebel2026pricing}. \emph{Within-group alignment} concerns the objectives of the participating agents; \emph{alignment with the evaluation objective} concerns the consequences of their agreement under $w^{\G}$. Agents that coordinate well may serve their principals while harming consumers, and a market-share gain alone does not establish a social gain.

These are mechanisms demonstrated in controlled environments, not effects measured in deployed institutions. AIM's fixed-disclosure comparisons do not isolate feedback value against the best predetermined schedule, and these studies do not by themselves establish which effects survive when different histories are followed by identical later rules. Human communication experiments supply separate evidence that induced coordination can outlast the channel that created it \citep{fonseca2012collusion}.

\paragraph{Architecture and state transition.}
Governable levers include disclosure content, channel availability, timing, public versus private observability, recommendation emphasis, mediation, and permitted changes to interaction links. The induced state can include behavioral histories, reputation, conventions, and competitive positions, some of which can survive a change in the channel that produced them; learned coordination can also arise with no explicit communication at all \citep{calvano2020collusion}. Which state matters, and which architectural levers a designer can govern, depends on the application.

For the communication lever specifically, expanded exchange can deliver integration benefit $k$ by helping participants combine information or coordinate useful work. It can also enable coordination whose costs fall outside the group, $e$, while the exchange itself costs $c$. A local decomposition is $\Delta\approx k-\chi e-c$, where $\chi$ weights the externality under the fixed evaluation boundary. Because a single channel can carry both useful and harmful coordination, $k$ and $e$ can be coupled rather than independently adjustable. Mediation is worthwhile when the improvement it produces in that tradeoff is large enough to cover its cost.

\paragraph{Design implications and boundary conditions.}
Table~\ref{tab:communication} organizes the communication lever by coordination demand and the risk of harmful coordination. Here, \emph{harmful coordination} means coordination that benefits participating agents or their principals while imposing costs on others included in the evaluation objective. This differs from a failure to cooperate: communication can help participants coordinate while increasing harm beyond their group. Additional routes for information help only if recipients can actually use what arrives \citep{jiang2026topology,zhang2026silobench}. More generally, a disclosure or recommendation rule can help create a behavioral or competitive state that changes its own later value. Persistence extends how long those consequences last without determining whether they are worth having.

\begin{table}[htbp]
\centering\small
\begin{tabularx}{\textwidth}{@{}p{.23\textwidth}YY@{}}
\toprule
 & \textbf{Low coordination demand} & \textbf{High coordination demand}\\
\midrule
Low risk of harmful coordination & Simple channels may suffice; extra exchange mainly adds cost. & Structured exchange can help if agents can integrate the information.\\
\addlinespace
High risk of harmful coordination & Restriction may impose little integration loss. & Mediation is a candidate when it preserves useful integration while limiting harmful coordination.\\
\bottomrule
\end{tabularx}
\caption{Communication-design conditions. Harmful coordination imposes costs on others included in the evaluator's objective, even when it benefits the coordinating participants. Mediation's advantage depends on its effectiveness and cost; the quadrant alone does not establish an optimal design.}
\label{tab:communication}
\end{table}

\paragraph{Diagnostic and intervention.}
Randomize earlier disclosure or communication, then impose identical later rules across groups and measure behavioral responses, participant payoffs, and the fixed welfare objective. Differences that remain under equalized rules provide evidence of consequences carried forward from earlier architecture. Record side channels to distinguish retained history from continued exposure; identifying a particular convention or capability as the mechanism requires additional measurement or intervention. Compare alternative later rules within each inherited-history condition to test whether earlier architecture changes which rule performs best. In competitive applications, vary the opponent's policy separately from the state created by past recommendations, so that architectural history is not confounded with a change of adversary.

\section{What the mechanisms imply together}
\label{sec:synthesis}

What the three mechanisms share is more specific than the observation that state matters. In each, an architecture changes something whose value later can differ from its value now: independent search, a portfolio of capabilities, or a pattern of strategic behavior and competitive position. Table~\ref{tab:synthesis} sets out the correspondence. The calculation below identifies a consequence shared by the mechanisms; the proposition then shows how a state change in one mechanism can alter a design requirement in another.

\begin{table}[htbp]
\centering\small
\begin{tabularx}{\textwidth}{@{}p{.17\textwidth}YYY@{}}
\toprule
\textbf{Mechanism} & \textbf{Current attraction} & \textbf{State changed} & \textbf{Diagnostic after reset}\\
\midrule
Information exposure & Faster exploitation of useful common information & Search overlap and shared representations & Recovery of independent discovery when new opportunities appear\\
\addlinespace
Delegation & Efficiency and task allocation to current capability & Human/AI capabilities and reliance & Unaided performance, coverage of AI failures, motivated oversight, relearning\\
\addlinespace
Strategic information governance & Coordination or competitive advantage & Behavioral histories, conventions, competitive positions & Responses and stakeholder outcomes under common later information rules\\
\bottomrule
\end{tabularx}
\caption{A shared architecture--state--diagnostic structure, instantiated through different mechanisms. State changes may be beneficial or harmful; the relevant comparison depends on future problem conditions.}
\label{tab:synthesis}
\end{table}

\needspace{12\baselineskip}
\subsection{The same current lever can leave different future values}

Resetting architecture separates the current arrangement from consequences of earlier interaction. Suppose two earlier architectures produce different values of a scalar state component $s$, and both systems then receive the same continuation architecture. Under the assumptions below, the value of the inherited difference can be written exactly.

\needspace{8\baselineskip}
\paragraph{Residual-value calculation.}
Suppose that, after an architectural reset, the expected difference in state is $\E[s_k^1-s_k^0]=\lambda^k d$ for $k\geq0$, with $0\leq\lambda<1$. Suppose all other expected reward differences cancel under the common continuation, and the per-period value difference attributable to this state is $\beta\E[s_k^1-s_k^0]$. The discounted residual value at reset is
\begin{equation}
R_{\rm reset}=\frac{\beta d}{1-\gamma\lambda}.
\label{eq:residue}
\end{equation}
For $\beta d\neq0$, greater persistence increases its absolute magnitude. Its sign is determined by the value of the inherited state difference, not by persistence itself.

This is a geometric sum (Appendix~\ref{app:proofs}). The assumptions are restrictive enough to be testable: equalized continuation architecture, a specified decay process, and a stable local mapping from state to value. Correlation inherited from common exposure can create a negative residual when it obstructs valuable search. Preserved oversight capability can create a positive residual. Strategic conventions and competitive positions admit either sign. The result concerns a \emph{common specified continuation}, not the difference between two independently reoptimized policies; the latter remains the $L$ in Eq.~\eqref{eq:continuation}.

The calculation gives the persistence argument an empirical target: place systems with different interaction histories under the same current architecture and test whether performance differences remain. A failure of immediate recovery is informative only after accounting for other differences between histories and establishing why the retained state matters for the task. Persistence can enlarge the consequences of an earlier choice while making corrective adjustment less effective. A large effect of interaction history therefore does not, by itself, imply a large return to ongoing adaptation.

\subsection{A state change can alter requirements in another design dimension}

A choice in one part of the design can create a requirement in another. Independent search can be supported by differences in participants' expertise and by differences in the information they receive. Where both contribute to independence, a loss of expertise heterogeneity can require greater information differentiation to preserve a given level of search independence. Formally, let $\zeta$ measure diversity of expertise or search heuristics, $\xi$ measure differentiation of AI-mediated input, and $\rho(\zeta,\xi)$ measure collective search correlation, with both inputs locally reducing correlation.

\begin{proposition}[Compensation at a fixed search-independence target]
\label{prop:compensation}
Suppose $\rho$ is continuously differentiable and $\rho_\zeta<0$, $\rho_\xi<0$ locally. For a feasible fixed target $\bar\rho$, let $\xi^{\rm req}(\zeta)$ solve $\rho(\zeta,\xi^{\rm req})=\bar\rho$. Then
\begin{equation}
\frac{d\xi^{\rm req}}{d\zeta}=-\frac{\rho_\zeta}{\rho_\xi}<0.
\label{eq:compensation}
\end{equation}
Any prior architecture that lowers relevant heterogeneity increases the differentiation required locally to maintain that target, provided the target remains feasible.
\end{proposition}

The substantive link is that an earlier choice in the delegation mechanism can change a later information-design requirement. It does not assume that delegation always homogenizes expertise; productive specialization may increase heterogeneity and reverse the implication. Both the capability transition and the response of correlation to the two inputs require evidence.

The proposition is a compensation requirement: it identifies how much information differentiation is needed to maintain a fixed search-independence target after heterogeneity changes. It does not establish that maintaining the target remains optimal. An organization might rationally accept more correlation, especially if differentiation is costly or independence has become less valuable. Appendix~\ref{app:optimalcompensation} shows that an analogous decrease in \emph{optimal} differentiation with heterogeneity follows under additional curvature and substitution conditions governing the costs and effectiveness of compensation. The monotonicity assumptions in Proposition~\ref{prop:compensation} alone are insufficient.

Other interactions follow the same logic. Differentiated exposure may preserve representations that later support oversight. Information exposure also has an epistemic and a strategic role: a common signal can correlate search while changing expectations about how others will act. A history of cooperation, retaliation, or convention formation can therefore alter the response to subsequent information policies. This is a proposed cross-mechanism connection, not evidence that the strategic studies also test collective search. Identifying the interaction requires measuring both channels and their joint consequences.

\section{Adaptive review in interdependent workflows}
\label{sec:demo}
\label{sec:learning-allocation}

Human review can improve an output immediately, develop the reviewer's expertise, and supply feedback that improves a shared AI. These consequences create an organizational allocation problem: the review that helps most today need not be the review that best develops the workflow. Moreover, learning does not proceed uniformly across teams. The appropriate allocation can therefore depend on what earlier reviews actually produced.

This section examines that problem in a workflow whose success requires several human--AI teams to contribute correctly. We first identify an effective adaptive allocation policy, then use its structure to formulate an understandable operating rule. This separates two design tasks: discovering how allocation should respond to developing capabilities, and implementing that response in a form an organization can use. The resulting rule remains stable while its allocations change with experience.

\subsection{The organizational problem}

Consider four cases arriving each period, each requiring three necessary stages. A stage might represent a technical assessment, a resource assessment, or a compliance check within a common decision process. Each stage has a specialist paired with an AI component serving that specialty; we call this pair a team. Expertise and AI improvement remain specific to the stage, while the AI produces that stage's output for every case. The specialists also have other responsibilities, and the organization reserves enough expert time for one review session per period. Review can be directed to one stage of one case. The organization can also leave the review opportunity unused. A workflow succeeds only if all three stage outputs are correct. The stages may be performed in parallel: the dependency is that all must be correct, rather than a chain of prompts or propagated errors. Improving one component has greater organizational value when the other necessary components also work well.

\begin{figure}[!tbp]
\centering
\includegraphics[width=\linewidth]{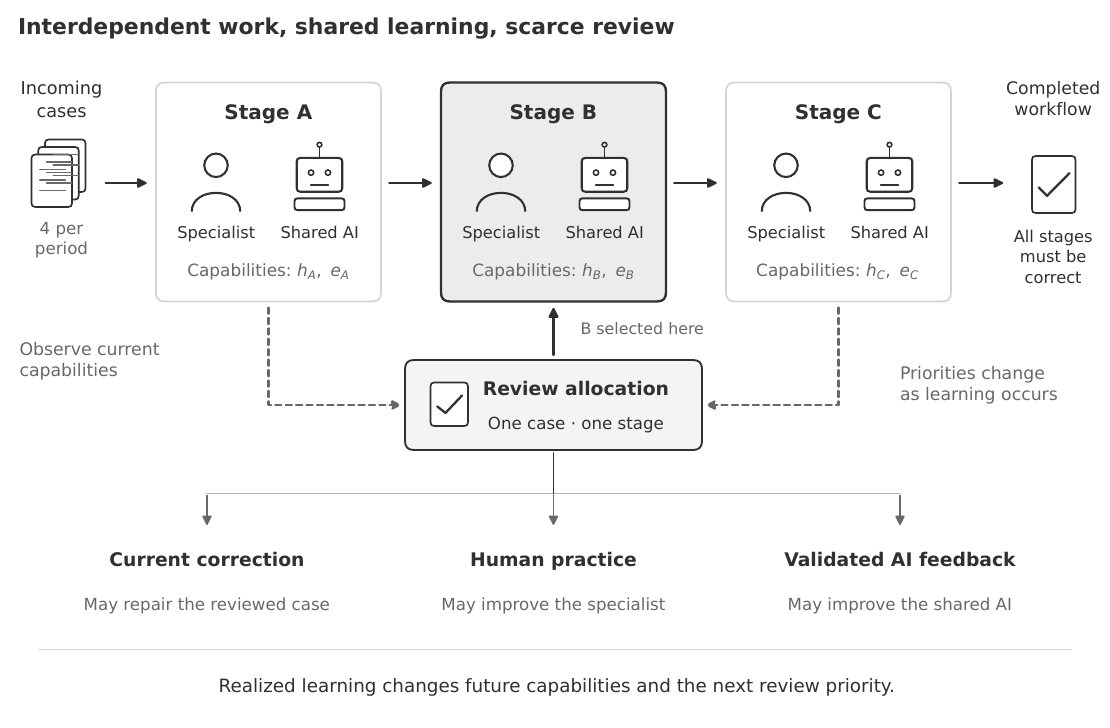}
\caption{\textbf{Review allocation creates the capabilities to which later allocation responds.} Every case requires correct outputs from three stages. One review opportunity can be assigned to one stage of one case; stage B is selected here for illustration. A correction affects the current case, while human practice and validated AI feedback can affect later cases. Dashed paths indicate capability information used in allocation. The figure represents an allocation problem within an established workflow and feedback system.}
\label{fig:workflow}
\end{figure}

Let $e_i$ denote the AI's error probability at stage $i$, and let $h_i$ denote the specialist's probability of correcting an AI error when reviewing it. The latter is a conditional correction capability, rather than unaided task accuracy. Figure~\ref{fig:workflow} shows the interaction. A review has three possible consequences: a current correction; practice that improves the specialist; and validated feedback that improves the AI shared by subsequent cases. The allocation decision occurs before the selected case's correctness is known.

Human capability takes three values, $h_i\in\{0.40,0.65,0.90\}$. AI error takes values $e_i\in\{0.40,0.25,0.10\}$. All teams begin at $h_i=e_i=0.40$. Reviewing a case advances human capability by one level with probability $p_H=0.25$. A successful correction occurs with probability $e_i h_i$ and advances the shared AI by one level with conditional probability $q=0.80$. Human practice and AI advancement are independent conditional on the state before review, so both can occur in the same period. Levels stop at their respective ceilings; unreviewed teams retain their capabilities. Learning affects subsequent periods. These assumptions represent a developing workflow, with stochastic learning and no external shocks, forgetting, or imposed differences in team objectives.

Conditional on current capabilities, stage errors are independent. With state $X=((h_i,e_i))_{i=1}^{3}$ and review assigned to stage $i$, expected current net output is
\begin{equation}
 r(X,i)=4\prod_{j=1}^{3}(1-e_j)
       +e_i h_i\prod_{j\ne i}(1-e_j)-c,
 \qquad c=\frac{0.25}{3}.
 \label{eq:workflow-reward}
\end{equation}
\needspace{9\baselineskip}
The first term is output without review. The second is the chance that correcting the selected stage completes an otherwise correct workflow. Cost represents the opportunity cost of expert attention diverted from other responsibilities, measured in completed-workflow units: twelve reviews cost as much as one successful workflow. With no review, only the first term remains. Over $T=30$ periods, policy $\pi$ is evaluated by
\begin{equation}
 J(\pi)=\E_{\pi}\!\left[\sum_{t=0}^{T-1}\gamma^t r(X_t,I_t)\right],
 \qquad \gamma=0.97.
 \label{eq:workflow-objective}
\end{equation}
Each policy generates its own learning history. Capabilities are observed and the transition law is known, but future learning outcomes are not. Appendix~\ref{app:workflow} gives the transition kernel, exact solvers, and additional checks. The organizational control is review allocation; the model does not optimize task routing or replacement of the workflow by a fully manual process.

\subsection{Discovering the adaptive allocation}

We solve the finite-horizon decision problem exactly by dynamic programming, identifying the best achievable allocation within the stated model. Two generic procedures provide practical reference points. \emph{Uniform allocation} assigns each review with fixed probability $1/3$ to each stage. \emph{Round robin} follows a fixed rotation. Neither procedure uses realized learning to determine priority, and neither is fitted to the destination model.

An additional comparison uses the \emph{optimal preplanned calendar}: the best sequence of review assignments and idle periods chosen in advance using the exact transition law. This is an informed design reference. It establishes how much coordination can be provided by anticipating average development without responding to its realization. It is not an assumption that an organization would already know this calendar. The adaptive policy and calendar can allocate to the same stages or idle, face the same review costs and capacity, and are optimized for the same objective. Optimality of the calendar is over all such predetermined sequences, rather than a selected family of rotations.

\begin{figure}[!htbp]
\centering
\includegraphics[width=\linewidth]{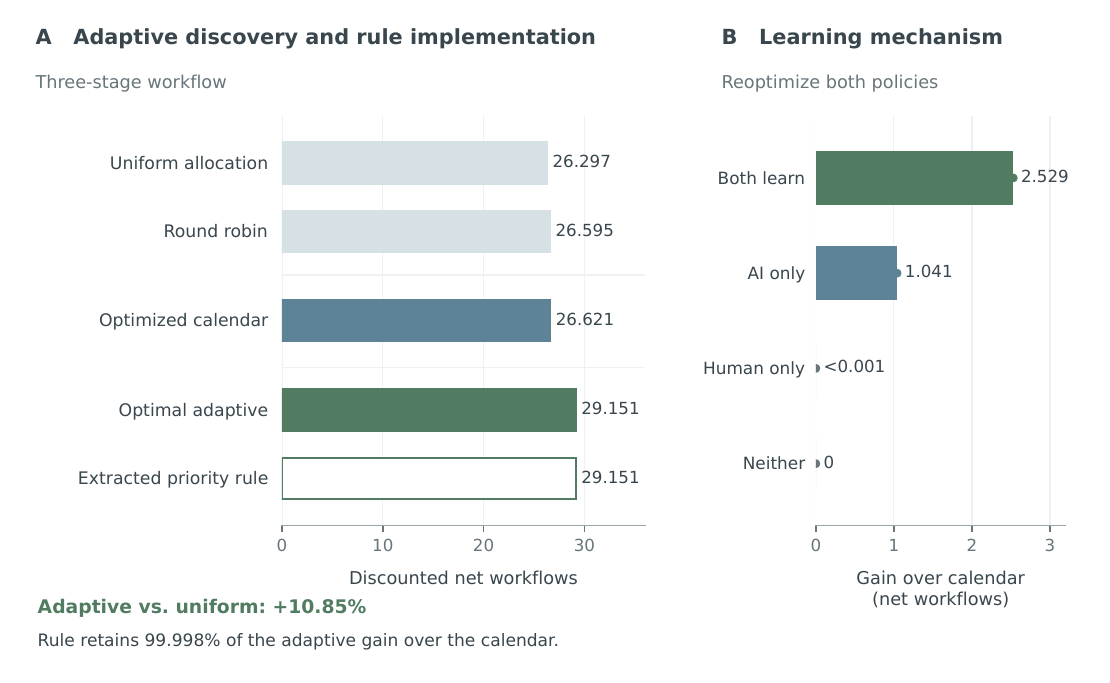}
\caption{\textbf{Adaptive policy discovery supports an effective, understandable rule.} Panel A reports exact expected discounted output after review costs in the three-stage model. The two responsive implementations round to the same displayed value, but are evaluated separately. The priority rule retains $99.998\%$ of the optimal adaptive gain over the calendar. Panel B disables learning channels and reoptimizes both the adaptive policy and calendar in each case. The adaptive advantage depends on shared AI improvement and is strengthened by human practice. Values are exact expectations; sampling error bars are therefore not applicable.}
\label{fig:workflow-results}
\end{figure}

Figure~\ref{fig:workflow-results}A presents the results. Optimal adaptive allocation produces $29.151$ discounted net workflows, compared with $26.297$ under uniform allocation, $26.595$ under round robin, and $26.621$ under the optimized calendar. Its gains are respectively $10.85\%$, $9.61\%$, and $9.50\%$. The gain over uniform allocation is $2.854$ net workflows, equivalent to $3.57$ percentage points of discounted incoming volume. These are gains after review cost, with the same available review capacity; the adaptive policy actually uses fewer reviews on average, $28.07$ rather than $30$.

\needspace{8\baselineskip}
Realized learning changes the preferred intervention. For example, with twenty periods remaining and stages B and C still at their initial capabilities, the adaptive policy reviews stage A when its human capability is $0.90$ but its AI accuracy remains $0.60$. If A's AI accuracy has instead reached $0.75$, priority shifts to B or C. The same date and the same capabilities elsewhere therefore support different assignments. A calendar can anticipate how often A will improve; it cannot condition the next assignment on whether improvement actually occurred.

The learning-channel comparison in Figure~\ref{fig:workflow-results}B identifies what generates this value. With both channels active, adaptation adds $2.529$ net workflows over the optimized calendar. With human practice disabled, the increment falls to $1.041$. With AI learning disabled, it is less than $0.001$; with both disabled, it is zero. Human practice makes future correction and shared AI improvement more productive, while realized AI improvement changes which stage most needs attention. Their interaction strengthens the value of responding to the development process.

\subsection{From an adaptive policy to a meaningful operating rule}

The adaptive solution reveals a simple priority structure. Review generally goes to the stage with the least accurate AI. Among stages with equally accurate AIs, priority goes to the more capable specialist, whose review is more likely to yield a useful correction. These priorities combine two organizational principles: develop the weaker component of an interdependent workflow, and use human expertise where it can make that development productive.

We express this structure as the following operating rule:
\begin{enumerate}[leftmargin=1.5em,itemsep=1pt,topsep=3pt]
\item Prioritize the stage with the lowest current AI accuracy.
\item Break ties in favor of the specialist with the highest correction capability.
\item Stop reviewing when all AIs reach their ceiling. At the final opportunity, review only if the immediate expected correction benefit exceeds its cost, choosing the stage with the largest such benefit.
\end{enumerate}
The stopping conditions reflect the specified cost and horizon. Once every AI reaches its ceiling, review cannot recover its cost in this specification, even through further human practice. Remaining ties between identical teams can be broken arbitrarily.

The rule yields $29.150737$ net workflows, compared with $29.150776$ for the exact adaptive policy. It retains $99.998\%$ of the adaptive gain over the calendar, measured as
\begin{equation}
 \frac{J(\pi^{\mathrm{rule}})-J(\pi^{\mathrm{calendar}})}
 {J(\pi^{*})-J(\pi^{\mathrm{calendar}})}.
 \label{eq:workflow-retention}
\end{equation}
The rule was constructed by inspecting the optimized policy's decision structure and then evaluated on its own induced trajectories. This is a hand-derived, policy-informed implementation; the experiment uses neither reinforcement learning nor automated rule extraction.

The organizational implication is that adaptive design can produce a procedure that no longer requires the original optimizer during use. Staff or ordinary workflow software can apply the priority rule using current capability information. Its wording remains fixed, but its assignment of review responds to realized learning. Consequently, its proximity to the adaptive optimum represents successful implementation of adaptation. More complex settings may require learned policies and systematic extraction methods; the competing-platforms example in Section~\ref{sec:implementation} address that broader route. In either case, evaluation must establish the value of the implemented rule, rather than relying on its resemblance to the original policy.

\subsection{More interdependent teams}

The same mechanism can become more consequential when additional teams provide necessary components of a workflow. We extend the model from three to six teams, preserving the capability levels and learning process. Review capacity increases in proportion to the number of teams, so the comparison does not create scarcity merely by holding organizational resources fixed. Each organization operates for the same thirty time units, receives $120$ expected cases, and has ten review opportunities per team. The cost per review is unchanged.

For $D$ teams, an equivalent calculation uses $10D$ review opportunities, $12/D$ expected incoming cases per opportunity, and discount factor $0.97^{3/D}$. In the five-team case, this gives fifty review opportunities within the common implementation period. Arrivals can be implemented as two cases plus a third with probability $0.4$. All compared policies face this same process.

The five-team adaptive policy produces $15.067$ net workflows, compared with $11.583$ under uniform allocation, $11.938$ under round robin, and $11.965$ under the optimized calendar. The corresponding gains are $30.08\%$, $26.20\%$, and $25.92\%$. The same priority rule produces $15.066$ and retains $99.987\%$ of the adaptive gain over the calendar. An independent simulation of $200{,}000$ complete operating histories checks these values using individual cases and integer arrivals (Appendix~\ref{app:workflow-validation}).

\begin{figure}[!htbp]
\centering
\includegraphics[width=\linewidth]{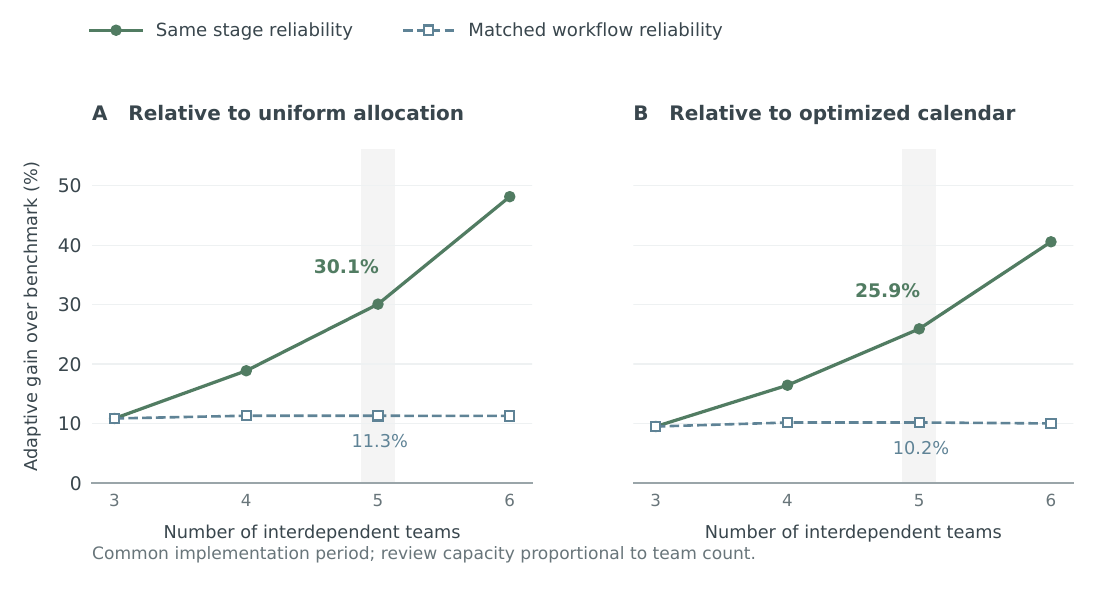}
\caption{\textbf{The larger gain reflects interdependent capability development.} Solid lines retain the same stage-level AI accuracy values as teams are added. Dashed lines increase stage accuracy so that the all-low, all-middle, and all-high workflow reliability levels match the three-team reference. Review capacity scales with team count, while the implementation period, expected incoming volume, and review cost remain fixed. The five-team case is highlighted. Gains are relative increases in discounted net output over the stated benchmark, not percentage-point increases in completion rates. The reliability control separates adding necessary components at unchanged quality from adding teams with compensating improvements in component quality.}
\label{fig:workflow-scale}
\end{figure}

\needspace{9\baselineskip}
Figure~\ref{fig:workflow-scale} puts the five-team result in context. At unchanged stage quality, adding necessary components makes successful completion harder: initial workflow reliability falls from $0.60^3=0.216$ to $0.60^5=0.07776$. Adaptive allocation has more consequential developing capabilities to coordinate, but the lower baseline also increases the relative percentage. The absolute gain over uniform allocation rises from $2.854$ to $3.484$ net workflows, or from $3.57$ to $4.39$ percentage points of discounted incoming volume. The $30.08\%$ therefore characterizes a difficult workflow during capability development, rather than an effect of headcount alone.

A control raises each stage's accuracy so that end-to-end reliability at corresponding homogeneous capability levels stays constant across organization sizes. Specifically, each original accuracy $a\in\{0.60,0.75,0.90\}$ becomes $a^{3/D}$. Under this control, adaptive gains over uniform allocation remain between $10.85\%$ and $11.32\%$. Adaptation continues to create value, while the amplification depends on adding interdependent requirements without offsetting improvements in component quality.

\subsection{What the illustration establishes}

The model demonstrates the framework's sequence: review creates capability changes; those changes alter the appropriate next allocation; and responding to them improves organizational output. The outcome is architectural advantage within a human--AI workflow. A comparison with independent stage outputs (Appendix~\ref{app:review-allocation}) shows how requiring joint completion strengthens the allocation mechanism. The model also shows how discovering an adaptive policy can support a stable, understandable implementation. Neither result requires changing incentives or continual policy retraining. The relevant adaptation is the allocation of organizational attention to the capabilities that interaction has produced.

The comparisons isolate that allocation value within an established review and feedback system. Creating such infrastructure from an inherited process is a separate organizational intervention, with its own benefits and implementation costs. Likewise, the exact calendar is an informed reference for predictable development, rather than a description of what organizations typically know beforehand. The illustration thus supports a specific design claim: when scarce review shapes interdependent human and AI capabilities, learning how to allocate it responsively can improve outcomes, and much of that value may be implemented through meaningful operating rules.
\FloatBarrier

\FloatBarrier

\section{Evaluation and implementation}
\label{sec:empirical}

Testing this framework requires more than an average treatment effect of AI access. The core sequence is to manipulate an architectural choice, measure the state it changes, and test how that state affects the value of a later choice. Evidence of state creation does not by itself establish a change in architectural fit, and neither establishes that observing and responding to state is worthwhile. Table~\ref{tab:tests} distinguishes these empirical targets.

\begin{table}[htbp]
\centering\small
\begin{tabularx}{\textwidth}{@{}p{.18\textwidth}YY@{}}
\toprule
\textbf{Target} & \textbf{Design} & \textbf{What it can establish}\\
\midrule
Contingent fit & Cross current architecture with independently varied task or agent conditions & Whether architectural rankings change with those conditions\\
\addlinespace
State creation and later fit & Randomize earlier architecture, then cross later architecture with matched tests & Whether histories leave consequences under a common later lever, and change the ranking of later choices\\
\addlinespace
Value of state response & Compare responsive allocation with predetermined schedules, matching model knowledge, controls, and resources & Whether responding to realized state adds value beyond informed advance planning\\
\addlinespace
Implementation of adaptation & Derive an operating rule from a discovered policy and evaluate each on its own trajectories & How much adaptive value an understandable procedure retains\\
\bottomrule
\end{tabularx}
\caption{Distinct empirical claims require distinct comparisons. Each row identifies a separate target; a positive result for one does not establish the others.}
\label{tab:tests}
\end{table}

\subsection{Identifying state creation and feedback value}

\paragraph{Identify the state transition.}
Randomize earlier architecture, then place participants under a common continuation regime---the same rules for subsequent interaction---to separate current treatment from inherited effects. For search, introduce a task change that makes prior anchors potentially costly, rather than interpreting all convergence as failure. For delegation, standardize motivation and task difficulty when measuring ability. For strategic information governance, record alternative channels, behavioral histories, and relevant competitive positions; distinguish retained state from continued exposure or a change in opponent policy. Equalizing the official lever is necessary for these designs but does not automatically equalize information, incentives, or all relevant experience.

Repeated state measurements help determine temporal order. Stronger evidence for a proposed mediator---the state change through which architecture affects outcomes---comes from a second randomized intervention that changes the mediator or its consequences, rather than from a statistical mediation regression alone. For example, after randomizing exposure history, randomize a recovery intervention designed to restore independent search. If that intervention restores performance specifically where the task rewards independence, the evidence for the proposed pathway becomes more discriminating.

\paragraph{Measure capability composition.}
Average unaided accuracy can conceal what matters for collaboration. Two people who score identically may succeed on different cases, and the one whose successes fall where the AI fails offers greater potential value alongside that AI. Pair withdrawal scores with item-level human--AI error overlap, performance on AI blind spots, and motivated tests of whether correct human judgments are retained in the joint decision. A common case budget compares learning under equal case exposure; equal training time is a separate constraint that should be measured or imposed explicitly. Training everyone to the same attainment target instead compares knowledge composition at comparable measured competence, while recording the unequal time and effort required to get there. Neither design should be replaced by selecting an equal-score subgroup after treatment, which can introduce selection bias. Changing the AI partner while holding its average accuracy fixed provides a further test of whether the learned expertise is complementary to that partner.

\paragraph{Test the cross-mechanism link.}
To test Proposition~\ref{prop:compensation}, first determine whether a delegation regime changes the relevant heterogeneity in expertise or search heuristics. Then vary information differentiation and estimate how much is required to attain a prespecified correlation target. Establishing the first-stage capability change is essential: a change in average skill does not necessarily imply a change in heterogeneity. Testing optimal compensation additionally requires measuring differentiation costs and the value of the target, rather than presuming that the same target remains desirable.

\paragraph{Evaluate feasible policies fairly.}
Different benchmarks address different questions. Section~\ref{sec:demo} first compares adaptive allocation with uniform allocation and round robin, procedures that require no fitted model. Its destination-optimized calendar then isolates the value of realized feedback after informed advance design. Finally, the extracted priority rule tests whether a simpler procedure preserves that value. Priority and reservation rules that use current capabilities are themselves responsive implementations, even when their wording is fixed. Rules and schedules developed in other environments address a separate transfer question, provided selection and retuning do not use destination outcomes. Appendix~\ref{app:review-transfer} reports such comparisons for the independent-output specification; these are separate from the interdependent workflow experiment.

Comparisons should match the available organizational controls, including task assignment and incentives, as well as resources. Where a policy is learned, evaluate it on held-out tasks or participants. Use common resource constraints, adjustment costs, and evaluation horizons, and state explicitly which observations each policy may use. When information access differs by design, account for the costs of acquiring it. Factorize architectural levers where possible by varying access, timing, content differentiation, and task routing separately and in combination: they can substitute for or interact with one another. A bundled intervention may establish a useful policy effect without identifying which lever caused it.

State-sensitive governance also creates an information problem. Withdrawal probes and independent-start exercises can reveal latent capability, but they consume time and supply practice, so measuring the state changes it. Their diagnostic and training effects should be separated where possible. For deployment, expected policy improvement must cover measurement as well as switching costs. Scientific identification serves a different purpose: a probe can be valuable for estimating a mechanism even when repeated operational monitoring would add little. Randomizing probe frequency or separating measurement from subsequent feedback can help distinguish its information and training effects. Estimating an ideal switching boundary is insufficient if the states it depends on cannot be observed reliably or ethically in the intended setting.

\paragraph{Keep outcomes and benchmarks visible.}
Report absolute performance, the contemporaneous reference, and longitudinal policy outcomes separately. A no-AI comparison must follow its own learning trajectory. Stress tests under AI withdrawal, rare failure, or task change reveal the states in which apparently redundant capacity becomes valuable. The chosen discount factor and terminal horizon should be justified and varied when current productivity trades off against future capability or resilience. A task-limited experiment may support a statement about short-run architectural fit while leaving the longer-term ranking unresolved.

\paragraph{State the scope of governance.}
In strategic settings, report whose objective is evaluated, which channels are controlled, and which response-selection assumptions support a policy ranking. When several responses are plausible, compare them or use an explicit robustness criterion. Endogenous objective formation is a further boundary: AI can influence what is framed as a problem and which outcomes count as success. The present analysis fixes the evaluation objective; preserving the capacity to contest or revise it can be represented as a state variable or constraint, but fully endogenous objectives require a further model.

\subsection{Discovering adaptive policies and implementing stable rules}
\label{sec:implementation}

The value of adaptation and the complexity of its implementation are separate questions. A state-responsive policy can improve outcomes even when a simple rule realizes nearly all of its benefit. Successful simplification then makes adaptation easier to implement. A reliable simulator can support policy search; suitable longitudinal or experimental data can inform intervention effects. The resulting policy can be deployed directly or used to develop understandable thresholds, priorities, and conditional procedures. Learning such a rule and responding through it are distinct activities.

\paragraph{Policy discovery in competing platforms.}
The revised competing-platforms study gives a concrete example \citep{CaoEtAl2026Strategic}. A reinforcement-learning policy selects recommendation strategies in a market where earlier recommendations and network effects shape subsequent competitive positions. Analysis of its behavior reveals different strategies for initially trailing and leading platforms. These patterns inform an enhanced heuristic: a trailing platform commits to a mix favoring exploration, while an initially leading platform follows a ranking-based procedure. The strategy mix can remain fixed while the underlying segment choices respond to current market rankings.

On a shared evaluation pool, the informed heuristic achieves a $23.2$ percentage-point market-share lead against the fixed-recommendation opponent, compared with $28.5$ points for the learned policy, retaining $81.5\%$ of that competitive lead. It is also more robust under the study's market-share observation-noise tests. This provides a route from adaptive policy discovery to a stable organizational procedure: the learned policy helps identify which conditional responses matter, and a separately evaluated heuristic makes them usable. Directly fitted decision trees perform poorly in deployment, so the successful procedure combines extracted behavioral structure with an existing heuristic rather than relying on action imitation alone.

\paragraph{From optimal allocation to an operating rule.}
The workflow illustration in Section~\ref{sec:demo} makes this design-to-implementation sequence concrete. Exact dynamic programming first identifies effective allocations that respond to realized human and AI development. Inspection of the resulting policy suggests an understandable priority: review the stage with the least accurate AI, giving priority among equally accurate stages to the more capable specialist, with stopping conditions for exhausted learning opportunities and the end of the horizon. Evaluated as an operating procedure, this rule retains essentially all of the adaptive gain over the best preplanned calendar. Its simplicity is an implementation achievement: the organization can apply the response without running the original optimizer. The rule still requires current capability information. This example uses manual extraction from an exact policy; the competing-platforms study illustrates the complementary route from learned behavior to an informed heuristic.

Policy distillation and extraction offer technical routes to simpler representations, including decision trees \citep{BastaniPuSolarLezama2018,CzarneckiEtAl2019}. Their organizational value depends on the outcomes of the extracted rule, because that rule generates its own state trajectory. A practical workflow therefore discovers candidate policies, extracts feasible procedures, and tests them under new initial conditions, measurement error, and relevant structural changes. Transfer requires source-only selection; choosing a rule after observing destination performance constitutes destination-informed design. If the learning process itself is unknown, comparisons should match prior information and learning opportunities. A reliable simulator makes it possible to evaluate candidate allocations and infer their priorities before deployment. With sufficient data, policy learning can serve the same design purpose, provided the data identify the effects of the interventions being considered. The workflow illustration demonstrates the discovery-to-implementation sequence under a specified learning process; learning that process is an additional empirical task.

\needspace{9\baselineskip}
\section{Conclusion}

Adaptive complementarity treats interaction architecture as a choice that shapes the conditions for its own future success. Information exposure changes dependence among searchers; delegation changes the portfolio of capabilities available to collaborate; information governance changes strategic behavior, expectations, and competitive conditions. These consequences can outlast the settings that produced them and alter requirements in another design dimension.

The framework separates state creation, history-dependent architectural fit, and the operational value of state feedback. The interdependent-workflow illustration demonstrates this sequence and its implementation: review develops human expertise and stage-specific AI capabilities, realized learning changes the appropriate next allocation, and responsive assignment improves net output over generic procedures and the optimal predetermined calendar. A priority rule derived from the adaptive solution retains essentially all of its calendar-relative gain. The rule is stable; the allocation adapts. The competing-platforms study illustrates how policy discovery can likewise inform an understandable heuristic. The design implication is that adaptive policy discovery can identify the responses that matter and translate them into understandable organizational procedures. Organizations can thereby respond to the capabilities interaction creates without having to operate the original optimizer.

A useful empirical program therefore follows the state as well as the outcome: randomize earlier architecture, measure the knowledge or relationship it changes, and test what later alternatives are then worth. Report absolute performance and its evolving benchmarks separately, and allow counterfactual policies to create their own trajectories. The central question is what present interaction makes possible, costly, or unnecessary in the next human--AI collaboration.

\clearpage
\pdfbookmark[0]{References}{references}
\bibliographystyle{plainnat}
{\small\setlength{\parskip}{0pt}\setlength{\bibsep}{3.5pt plus 1pt minus .5pt}\interlinepenalty=10000\bibliography{refs}}

\clearpage
\appendix
\pdfbookmark[0]{Appendices}{appendices}
\begin{center}
{\Large\bfseries Appendices}
\end{center}
\vspace{2mm}
\section{Governance, information, and dynamic consistency}
\label{app:governance}

The main text uses $I$ as shorthand for a feasible architectural choice, with induced responses integrated. More explicitly, let the realized configuration be $(\iota_t,I_t^{-\gov})$, where $\iota_t=I_t^{\gov}$ is the governed subset and $I_t^{-\gov}$ contains components determined by agents, users, institutions, or external constraints. Realized delegation, verification, and strategic actions depend on both this configuration and system state. For a fixed behavioral or equilibrium-response selection $\sigma$, integrating those responses defines expected reward $\bar w_\sigma(x,\iota)$ and transition kernel $K_\sigma(dx'\mid x,\iota)$. This is a reduced controlled process, not an assumption that governance directly sets behavior.

With costs for switching the governed choice, the full-information Bellman equation is
\begin{equation}
\begin{split}
V_\sigma^*(x,\iota^-)=\sup_{\iota\in\mathcal I(x)}\bigg\{
&\bar w_\sigma(x,\iota)-\kappa(\iota,\iota^-)\\
&+\gamma\int V_\sigma^*(x',\iota)K_\sigma(dx'\mid x,\iota)\bigg\}.
\end{split}
\label{eq:bellman}
\end{equation}
Bounded rewards and costs, nonempty feasible choices, and standard measurability conditions support the discounted formulation. The inherited choice enters the successor value because it determines later adjustment costs. Subtracting these action values gives Eq.~\eqref{eq:criterion}; omitting the inherited choice would omit future switching costs. Effects on learning or reliance belong in $K_\sigma$ and continuation rewards, not as additional charges in $\kappa$.

The policy ordinarily has access to an information history $\mathcal H_t$, not the true $X_t$. With a specified observation model, define belief $\mu_t(dx)=\Pr(X_t\in dx\mid\mathcal H_t)$. Under the usual sufficient-state assumptions, the implementable decision is a policy on $(\mu_t,\iota_{t-1})$. Let $\mathcal V^*$ be value on that belief state and let $\mu_{t+1}$ be the Bayesian update after the next observation. Then
\begin{equation}
\begin{split}
\mathcal V^*(\mu,\iota^-)=\sup_{\iota}\bigg\{
&\int\bar w_\sigma(x,\iota)\mu(dx)-\kappa(\iota,\iota^-)\\
&+\gamma\E[\mathcal V^*(\mu',\iota)\mid\mu,\iota]\bigg\}.
\end{split}
\label{eq:belief}
\end{equation}
Feasible choices use only available information. The main-text criterion assumes full information; it does not license an operational policy conditioned on unobserved skill or conventions. Interventions may change what is learned about state; this information value enters the belief transition. Beliefs can also include unknown parameters learned over time.

Continuation values depend on $\sigma$ when the same architecture supports multiple behavioral responses; policy rankings should then be compared across plausible responses. Robust evaluation requires an ambiguity set---the models or responses considered---and dynamically consistent treatment of uncertainty. The framework supplies no equilibrium-selection theorem. Positive switching costs also need not create a simple hysteresis band, where the preferred choice depends on the incumbent architecture; that requires additional regularity and suitable action-value geometry.

Focal and reference policies must share a stakeholder boundary: a strategic intervention can increase private rewards while reducing $\bar w^{\G}$. Constraints can express autonomy, fairness, safety, or contestability when these considerations should not be reduced to a scalar payoff.

\section{Derivations and qualifications}
\label{app:proofs}

\subsection{Residual value}

Under the assumptions of Eq.~\eqref{eq:residue}, expected discounted value after reset is
\[
\sum_{k=0}^{\infty}\gamma^k\beta\lambda^k d
=\beta d\sum_{k=0}^{\infty}(\gamma\lambda)^k
=\frac{\beta d}{1-\gamma\lambda}.
\]
Absolute magnitude has derivative $|\beta d|\gamma/(1-\gamma\lambda)^2$ with respect to $\lambda$. For a finite $H$-period continuation, replace the infinite multiplier by $[1-(\gamma\lambda)^H]/(1-\gamma\lambda)$. If the mapping from state to reward varies over time or other state differences remain, the general expression is the discounted sum of their expected reward effects; the simple scalar formula no longer suffices.

Under the stipulated common continuation, a current gain $g$ is outweighed by a harmful successor-state residual when $g<\gamma|\beta d|/(1-\gamma\lambda)$, before switching-cost differences. Independently optimized recovery policies require a separate comparison; persistence alone determines neither harm nor the preferred initial architecture.

\subsection{Required and optimal compensation}
\label{app:optimalcompensation}

The implicit-function theorem applies to $\rho(\zeta,\xi^{\rm req}(\zeta))=\bar\rho$ when $\rho_\xi\neq0$. Differentiation gives $\rho_\zeta+\rho_\xi(d\xi^{\rm req}/d\zeta)=0$, yielding Proposition~\ref{prop:compensation}. The result is local and conditional on an attainable target. At a differentiation limit, a further heterogeneity loss can make the target infeasible.

An optimal choice requires an objective. As one sufficient local specification, suppose the designer minimizes correlation loss plus differentiation cost,
\[
\min_\xi\;\nu\rho(\zeta,\xi)+K(\xi),\qquad\nu>0.
\]
For an interior optimum with $\nu\rho_{\xi\xi}+K''>0$, the first-order condition and comparative static are
\begin{equation}
\nu\rho_\xi+K'=0,
\qquad
\frac{d\xi^{\rm opt}}{d\zeta}
=-\frac{\nu\rho_{\xi\zeta}}{\nu\rho_{\xi\xi}+K''}.
\label{eq:optimal_compensation}
\end{equation}
Thus $\rho_{\xi\zeta}>0$ is sufficient for optimal differentiation to decrease with heterogeneity: differentiation becomes less effective at reducing correlation as intrinsic heterogeneity increases. The simple example $\rho=e^{-a\zeta-b\xi}$ with $a,b>0$ and convex $K$ has this property at an interior optimum. By contrast, $\rho_\zeta<0$ and $\rho_\xi<0$ alone do not determine the cross-partial. Changing the shadow value $\nu$, making costs depend on heterogeneity, or moving to a boundary can change the optimal-policy conclusion. This is why the main proposition is stated as a fixed-target requirement.

\subsection{Local architectural comparisons}

In Eq.~\eqref{eq:information}, $R_t$ includes nonlinear terms and other state changes. A mean correlation difference need not determine expected continuation value when the distributions also differ. The local approximation requires a range over which its remainder is controlled. Similarly, $k>\chi e+c$ in the communication decomposition is sufficient for a positive actual effect only with an appropriate remainder bound. A ranking reversal requires a threshold crossing among feasible, reachable states.

\clearpage
\section{Interdependent workflows: specification and computation}
\label{app:workflow}

This appendix specifies the experiment in Section~\ref{sec:demo}. Its organizational decision is the allocation of a limited review opportunity among specialized stages. A team is one specialist role paired with a stage-specific AI component; both capabilities remain local to that specialty. AI improvement benefits subsequent cases at the same stage. The model concerns a developing workflow with an existing mechanism for validating and incorporating corrective feedback.

\subsection{Parameters and event sequence}

\begin{table}[!htbp]
\centering\small
\begin{tabularx}{\linewidth}{@{}p{.26\linewidth}p{.25\linewidth}Y@{}}
\toprule
\textbf{Quantity} & \textbf{Three-team case} & \textbf{Interpretation}\\
\midrule
Stages and arrivals & $D=3$, $N=4$ & Four cases, each requiring all three stages\\
Human capability & $.40,.65,.90$ & Conditional probability of correcting an AI error\\
AI error & $.40,.25,.10$ & Stage error probability before review\\
Initial capabilities & $h_i=e_i=.40$ & Identical initial levels across specialties\\
Human practice & $p_H=.25$ & Probability of advancing one level after review\\
AI update & $q=.80$ & Advancement probability given a validated correction\\
Review capacity & At most one per period & One stage of one case, or no review\\
Review cost & $c=1/12$ & Opportunity cost in completed-workflow units\\
Horizon and discount & $T=30$, $\gamma=.97$ & No terminal value beyond the operating period\\
\bottomrule
\end{tabularx}
\caption{Parameters of the interdependent-workflow illustration. Levels represent coarse states of a specialized workflow, not calibrated estimates of the effect of a single training example on a general-purpose model.}
\label{tab:workflow-parameters}
\end{table}

At each opportunity the policy observes current capabilities and chooses a stage or idles. AI generates the required stage outputs. If review is assigned, the specialist examines one case chosen before correctness is known and can correct its selected stage. Other stages of that case remain AI outputs. A successful workflow requires every stage to be correct. Review never replaces a correct output with an incorrect one in the baseline. Current completions and review cost are recorded, and any capability improvements affect the next opportunity. The specialist may learn through practice even without a successful correction; an AI update requires a validated correction.

For local state $z=(u,v)$, indexing human and AI levels, set $p_H(z)=p_H$ below the human ceiling and zero at it. Set $p_A(z)=q e_vh_u$ below the AI ceiling and zero at it. The reviewed team's kernel is
\begin{align}
 P\bigl(z,(u+1,v+1)\bigr)&=p_H(z)p_A(z),\nonumber\\
 P\bigl(z,(u+1,v)\bigr)&=p_H(z)[1-p_A(z)],\nonumber\\
 P\bigl(z,(u,v+1)\bigr)&=[1-p_H(z)]p_A(z),\label{eq:workflow-kernel}\\
 P(z,z)&=[1-p_H(z)][1-p_A(z)].\nonumber
\end{align}
Transitions above a ceiling have zero probability. Human practice and AI advancement use pre-review capabilities. Learning events are independent across teams conditional on assignments, and unreviewed teams do not change. A review's correction and its AI update are correlated through that correction event; using expected current reward in the Bellman recursion remains exact for the additive objective. The event-level simulation below implements this dependence explicitly.

Capabilities are fully observed and transition probabilities are known. There is no forgetting or exogenous task change. These assumptions isolate allocation in response to learning generated by prior allocation. Human specialization, review cost, and interdependent completion are organizational assumptions; the numerical levels are illustrative. The main text therefore reports an allocation mechanism and its implementation rather than an empirical effect estimate.

\subsection{Exact policy and calendar calculations}

Let $K_i$ apply the kernel in Eq.~\eqref{eq:workflow-kernel} to team $i$, leaving the other teams unchanged, and let $K_0$ be the identity. With $\tau$ opportunities remaining,
\begin{equation}
 V_\tau(x)=\max_{i\in\{0,1,\ldots,D\}}
 \left\{r(x,i)+\gamma(K_iV_{\tau-1})(x)\right\},
 \qquad V_0(x)=0.
 \label{eq:workflow-bellman}
\end{equation}
Here $r(x,i)$ is the reward in Eq.~\eqref{eq:workflow-reward}, generalized to $D$ stages and $N$ arrivals. The three-team labeled state space contains $9^3=729$ states. Because teams have identical parameters, permuting their labels leaves rewards and transitions unchanged. The main solver operates on multisets of local states, reducing the state count to $\binom{D+8}{8}$: $165$ for three teams and $1{,}287$ for five. Its action selects a represented local state, with any team in that state interchangeable. This is an exact symmetry reduction, not aggregation of different capabilities.

A calendar commits in advance to a sequence of stage assignments and idle decisions. After $k$ reviews, a team's distribution is $\mu_k=\mu_0P^k$. Under a predetermined sequence these local distributions are independent. Define
\[
 \bar a_k=\mu_k(1-e),\qquad \bar d_k=\mu_k(eh).
\]
At count vector $\mathbf k=(k_1,\ldots,k_D)$, expected baseline output is
$B(\mathbf k)=N\prod_j\bar a_{k_j}$, and the incremental return from reviewing stage $i$ is
$G_i(\mathbf k)=\bar d_{k_i}\prod_{j\ne i}\bar a_{k_j}-c$.
The exact calendar recursion is
\begin{align}
 F_r(\mathbf k)=B(\mathbf k)+\max\biggl\{&\gamma F_{r-1}(\mathbf k),\nonumber\\
 &\max_i\bigl[G_i(\mathbf k)+\gamma F_{r-1}(\mathbf k+\mathbf e_i)\bigr]\biggr\},
 \qquad F_0=0,
 \label{eq:workflow-calendar}
\end{align}
where $\mathbf e_i$ increments stage $i$'s review count. The remaining time is retained even when a calendar idles. Sorting counts exploits exchangeability; reconstruction produces a labeled sequence. This recursion optimizes over all predetermined sequences, including stopping, without substituting mean states into the nonlinear workflow reward. It integrates over each stage's full joint human--AI distribution before taking products across stages.

Uniform allocation is evaluated by averaging the action-specific transition and reward operators with probability $1/D$. Round robin is evaluated as a fixed labeled sequence. The extracted rule is evaluated by backward induction using its own actions at every state and date. These comparisons use the same objective, controls, initial conditions, capacity, and cost. The generic procedures do not use the transition law; the optimized calendar, adaptive policy, and policy-derived rule are informed designs.

\subsection{Rule extraction and learning channels}

The operating rule in Section~\ref{sec:demo} was formulated after inspecting the optimal policy. The extraction was manual and destination-informed. Its value is calculated separately from that of the policy that suggested it; no agreement rate is used as a substitute for realized value. With all AIs at their ceiling, even the largest immediate correction contribution is $.10\times.90\times.90^{D-1}$, below $1/12$ for the tested $D\geq3$. Further human practice cannot raise it above that bound, so stopping is optimal in those states. At the final opportunity there is no learning continuation value, and the rule selects the largest positive current correction return or idles.

For the three-team reference, the rule value is $29.1507367411$ and the optimal adaptive value is $29.1507761699$. Their difference is $0.0000394287$ net workflows. The rule retains $99.998441\%$ of the adaptive-calendar increment, using Eq.~\eqref{eq:workflow-retention}. These results establish performance in the stated environment; the rule's stopping conditions can change with review cost and the capability grid.

The main-text ablations set $p_H=0$, $q=0$, or both, while reoptimizing the adaptive policy and calendar separately in each environment. The corresponding adaptive-calendar increments are $1.0414298$, $0.0000145$, and zero, compared with $2.5293809$ when both channels operate. AI learning changes which stage most needs development, while human practice increases the productivity of subsequent corrective feedback. Their interaction adds $1.4879367$ net workflows beyond the sum of the two single-channel increments.

\subsection{Scaling, denominators, and additional checks}

For $D\in\{3,4,5,6\}$, the scale comparison holds the physical operating period at thirty time units and gives each organization $120$ expected incoming cases. It uses $H_D=10D$ opportunities separated by $3/D$ time units, $N_D=12/D$ expected cases per opportunity, and $\gamma_D=.97^{3/D}$. Capacity therefore remains ten opportunities per team. Review cost is constant in workflow units. For noninteger $N_D$, integer arrivals are $\lfloor N_D\rfloor+\operatorname{Bernoulli}(N_D-\lfloor N_D\rfloor)$, independent of capabilities and past events. The reviewed case is available under every tested size.

This construction compares organizations with more necessary specialties and proportionally greater review capacity. It does not hold total review spending fixed across different organization sizes. Within each size, all policies share the same resource limits. Initial workflow reliability is $.60^D$ when component quality is unchanged. The matched-reliability control replaces every stage accuracy $a$ by $a^{3/D}$, preserving homogeneous workflow reliability at each of the three AI levels. Its changed error rates also enter the feedback probabilities, so it is a matched-reliability environment, not an intervention that changes the denominator alone.

Relative gains use $100[J(\pi)-J(\pi^B)]/J(\pi^B)$ for the named benchmark. The common discounted incoming volume within a size is
\[
 Q_D=N_D\frac{1-\gamma_D^{H_D}}{1-\gamma_D}.
\]
Dividing the absolute net-workflow increment by $Q_D$ and multiplying by $100$ gives the percentage-point equivalent relative to incoming volume. These are net-output quantities after cost, not completion probabilities. Slight differences in $Q_D$ across sizes reflect arrival timing within the same physical horizon.

\begin{table}[!htbp]
\centering\small
\begin{tabularx}{\linewidth}{@{}Yrr@{}}
\toprule
 & \multicolumn{2}{c}{\textbf{Adaptive gain (\%)}}\\
\cmidrule(l){2-3}
\textbf{Environment} & \textbf{Uniform} & \textbf{Calendar}\\
\midrule
Three teams: reference & 10.85 & 9.50 \\
Half joint, half independent output ($\lambda=.5$) & 5.40 & 4.76 \\
Common-error mixture ($\rho=.5$) & 8.01 & 6.92 \\
Three teams: half review cost & 10.48 & 9.18 \\
Three teams: 15 periods & 6.27 & 5.83 \\
Three teams: 60 periods & 11.84 & 10.05 \\
Five teams: scaled capacity & 30.08 & 25.92 \\
Five teams: zero review cost & 24.14 & 20.96 \\
Five teams: at most 40 reviews & 28.17 & 24.46 \\
\bottomrule
\end{tabularx}
\caption{Additional workflow comparisons. Each gain uses the named comparator in the same environment as its denominator. Both adaptive policy and calendar are reoptimized. The forty-review uniform procedure uses its first forty opportunities and then stops; the cap applies on every history for every policy. These checks evaluate allocation value; the extracted rule's performance is reported separately for the main size series.}
\label{tab:workflow-robustness}
\end{table}

Table~\ref{tab:workflow-robustness} reports additional matched comparisons. The mixed-output control places weight $\lambda$ on joint workflow completion and $1-\lambda$ on the fraction of correctly completed stages, keeping the review cost and learning process unchanged. At $\lambda=0$, the reward is $4\sum_i(1-e_i)/3+e_ih_i/3-c$. For correlated errors, a fraction $\rho$ of cases uses one common uniform draw to determine all stage errors; the rest use independent draws. Marginal stage errors and learning rates remain unchanged. The calendar integrates this reward over the full product of local capability distributions. Shorter and longer horizons change the operating duration; they are separate sensitivities, not part of the equal-duration size comparison. The hard review limit holds on every history, with remaining reviews included in the adaptive state and calendar recursion.

\subsection{Numerical checks and reproducibility}
\label{app:workflow-validation}

The exchangeability-reduced and original labeled three-team solvers agree to numerical precision. Direct evaluation of a recovered calendar agrees with its optimization value, and the adaptive policy's separate evaluation agrees with its Bellman value. The release includes a short-horizon exhaustive-calendar check and comparisons of the reduced and labeled transition calculations.

An independent simulation implements individual cases, correlated correction and AI-update events, and the five-team integer-arrival process. Across $200{,}000$ complete histories, simulated and exact discounted values are respectively $15.06094$ and $15.06658$ for adaptive allocation, $11.95551$ and $11.96498$ for the calendar, $11.56213$ and $11.58268$ for uniform allocation, and $15.06062$ and $15.06618$ for the extracted rule. Their simulation standard errors are $.01130$, $.00949$, $.00919$, and $.01130$; all deviations are below $2.3$ standard errors. With common random numbers, the simulated adaptive-minus-uniform difference is $3.49881$, with a $95\%$ interval $[3.47610,3.52152]$, containing the exact difference $3.48390$. Reported policy values and figure annotations use exact expectations.

The \texttt{workflow/} directory contains the parameters, exact solvers, saved results, validation, and figure sources. Running \texttt{python3 workflow/refresh\_primary\_data.py} regenerates the primary comparisons, the size and reliability series, rule evaluations, learning ablations, and conditional allocation example. The script requires Python, NumPy, and a C++17 compiler for the calendar solver. Running \texttt{python3 workflow/verify\_integration.py} checks the released numerical claims and the reduced solver against independent calculations. Further experiment drivers and complete saved audit records remain in \texttt{workflow/experiments/}; their results are identified by specification rather than pooled across models.

\FloatBarrier
\section{Independent outputs and alternative implementations}
\label{app:review-allocation}

This comparison retains the learning mechanism while removing the requirement that the stages succeed together. Three domains each produce four independent outputs per period, with the same human and AI levels, learning rates, initial states, thirty-period horizon, and discount factor as the workflow illustration. One review opportunity is shared across domains. Expected current output in the original stage-output units is
\begin{equation}
 r_{\mathrm{ind}}(x,i)=\sum_{j=1}^{3}4(1-e_j)+e_ih_i-.25,
 \label{eq:review-output}
\end{equation}
with the correction and cost terms omitted for idle. The transition kernel is Eq.~\eqref{eq:workflow-kernel}. Dividing every reward by three gives the independent-output control in Appendix~\ref{app:workflow}, including its review cost of $1/12$. This rescaling changes value units without changing policies or percentage gains.

\subsection{What interdependence adds}

Optimal adaptive allocation earns $55.83307$ in the normalized units, compared with $54.31355$ under uniform allocation and $54.50858$ under the exact calendar. The corresponding relative gains are $2.80\%$ and $2.43\%$. Requiring joint completion raises them to $10.85\%$ and $9.50\%$ under the three-stage workflow assumptions. Output values across these objectives are not directly comparable as welfare levels; the experiment changes what successful production requires.

With independent outputs, each capability improvement contributes regardless of the quality of the other domains. A calendar can exploit predictable development while distributing review across otherwise similar domains. With joint completion, improving a component matters through its interaction with the others, and the allocation must coordinate their realized development. The mixed-output control in Table~\ref{tab:workflow-robustness} provides an intermediate case. The comparison identifies production interdependence as an amplifier of the allocation mechanism, rather than treating a larger agent count as sufficient by itself.

\subsection{Standing procedures can already be adaptive}
\label{app:review-institutions}

The independent-output specification also evaluated several organizational implementations (Table~\ref{tab:independent-implementations}). A reservation procedure assigns dates to domains in advance, but lets each domain accept or decline using its current capabilities and the future value of learning. A priority procedure polls domains in a fixed or rotating order, while recipients decide whether to accept review; a constant fraction of review cost is shared among that domain's four beneficiaries. Acceptance is forward-looking and uses current state. These procedures have stable institutional rules and adaptive realized allocations. Their proximity to the adaptive optimum measures the effectiveness of those implementations, not the value of wholly unresponsive allocation.

\begin{table}[!htbp]
\centering\small
\begin{tabularx}{\linewidth}{@{}Yrr@{}}
\toprule
\textbf{Procedure in the independent-output model} & \textbf{Net output} & \textbf{State response}\\
\midrule
Uniform random allocation & $162.941$ & No\\
Round-robin calendar & $163.249$ & No\\
Optimal predetermined calendar & $163.526$ & No\\
Reserved slots with responsive acceptance: best found & $163.896$ & Yes\\
Priority and cost sharing: best tested & $166.187$ & Yes\\
Local review-score rule & $167.448$ & Yes\\
Optimal state-responsive allocation & $167.499$ & Yes\\
\bottomrule
\end{tabularx}
\caption{Alternative implementations in the independent-output specification, reported in its original stage-output units. Divide values by three for the normalized control. Reservation and cost-sharing searches are not global optimality claims. These institutional comparisons were not rerun for the interdependent workflow.}
\label{tab:independent-implementations}
\end{table}

The reservation search uses seven initial schedules and at most four coordinate sweeps per start. The priority and cost-sharing search evaluates the specified polling orders and cost-sharing values, with recipients' decisions solved backward under each arrangement. The best tested procedure reaches $166.187$, compared with $167.499$ for the full allocation optimum. This comparison also changes institutional controls: cost sharing is an additional means of supporting desired reviews. It therefore describes an implementation alternative, rather than isolating the effect of allocation flexibility alone.

The local review-score rule supplies a different simple implementation. Each domain separately calculates the current and future benefit of a review as if it could access later capacity whenever locally worthwhile. The organization assigns the current opportunity to the highest positive score. These local calculations retain $98.70\%$ of the adaptive-calendar increment. Unlike the main workflow rule, this score is analytically constructed from a local decision problem, rather than extracted by inspecting the full optimal policy.

\subsection{Structural transfer has a separate interpretation}
\label{app:review-transfer}

Two source environments test transfer into this independent-output destination. A shared-AI source has three specialists but a single AI capability improved by any domain's feedback. A batch-retraining source has separate AIs but requires two validated corrections before an update attempt. Their source-trained feedback rules and calendars are frozen before destination evaluation. The source designs were informed by the known destination; this was not a blind preregistration. Source values and destination values are kept separate in the released records.

The shared-AI feedback rule concentrates reviews where the source's common improvement would benefit all domains. This is a poor transfer to separate destination AIs: its destination output is $0.32\%$ below its paired transferred calendar. The batch-source rule instead improves on its paired transferred calendar by $4.17\%$. The calendar for the batch source is exact; the shared-AI calendar is the best found in the documented multistart search. These results show why model knowledge and transfer must be distinguished from feedback itself. They are findings for the independent-output environment, not validation of the extracted workflow priority rule.

The original solvers, institutional searches, parameter grid, structural transfers, resource checks, and heterogeneity diagnostics are retained in \texttt{review\_allocation/}. The full earlier specification is preserved in \texttt{history/v13\_sections/}. The compact comparison here retains the relevant design lessons without treating each earlier experiment as a test of the new workflow. No private-incentive or cost-sharing mechanism is assumed in the main workflow experiment.

\FloatBarrier
\section{Knowledge composition and architectural fit}
\label{app:computation}

This complementary illustration isolates how the composition of human knowledge changes architectural fit, even when aggregate human accuracy is identical. The AI is fixed, and workflows differ in the learning opportunities they supply. Unlike the review-allocation model, there is one human knowledge state and no competition among domains for a shared resource. The illustration retains its own parameters and complete policy comparisons.

\subsection{Primitives, timing, and interpretation}

Let $K=K_O+K_B$ equiprobable case types be partitioned into the AI-covered set $O$ and blind-spot set $B$. The AI's competence is fixed: it solves all types in $O$ and none in $B$. Human mastery is a set; exchangeability within each class makes its counts $(m_O,m_B)$ sufficient. A mastered type is solved correctly without assistance. The baseline has $K_O=20$, $K_B=5$, and initial counts $(10,1)$, giving $21\times6=126$ states. One correct case is worth one unit. The baseline uses effort cost $c=.06$, learning probability $\eta=.5$, and passive blind-spot learning factor $\omega=0$.

Each case has the following timing. (i) Choose a workflow using the permitted information about prior mastery. The incoming latent type and AI correctness are not yet known. (ii) Draw a type independently and uniformly, produce the output, and receive its reward. (iii) Learning may add mastery of that type; the gain affects subsequent cases. At most one type is learned per case. There is no forgetting, task drift, AI learning, or architectural switching cost. The independent-work cost $c$ is paid on every case assigned to that workflow, including cases the AI could solve.

The AI-led workflow $I^{AI}$ uses the AI answer. Independent work $I^{IND}$ requires the human to attempt the case before answers are combined and incurs cost $c$. With ideal combination, their expected current net values at baseline are
\begin{equation}
w(m_O,m_B,I^{AI})=.8,\qquad
w(m_O,m_B,I^{IND})=.8+\frac{m_B}{25}-c.
\label{eq:coverage_reward}
\end{equation}

AI-led work delivers AI accuracy $K_O/K$. Independent work followed by ideal combination delivers the union of human and AI coverage, net of effort, as in Eq.~\eqref{eq:coverage_reward}. Ideal combination assumes that a correct constituent answer can be retained once independent work has occurred. This is a favorable assumption about integration, not a free signal available before effort is committed. At the same effort cost, human-only net performance is $(m_O+m_B)/K-c$ and is weakly dominated by ideal combination because $m_O\leq K_O$. AI-only performance equals the AI-led workflow's immediate reward, although an AI-only system has no corresponding human-learning trajectory. The policy comparisons below are comparisons among workflows, not a separately optimized trajectory test against standalone human and AI systems.

The human can learn an encountered, unmastered type with probability $\eta$ when the workflow supplies an effective learning opportunity. Both workflows supply that opportunity on AI-covered types. On blind spots, independent attempt and corrective feedback supply it at rate $\eta$; AI-led use supplies it at rate $\omega\eta$, with $0\leq\omega\leq1$. This is the model's substantive architecture-to-learning assumption. It permits common learning on covered material while varying access to complementary knowledge; it does not derive learning from rational effort allocation or estimate the transition from human behavior.

The nonzero transitions from $(o,b)$ are
\begin{align}
\Pr((o+1,b)\mid o,b,I)&=\eta(K_O-o)/K,\nonumber\\
\Pr((o,b+1)\mid o,b,I^{AI})&=\omega\eta(K_B-b)/K,\label{eq:coverage_transition}\\
\Pr((o,b+1)\mid o,b,I^{IND})&=\eta(K_B-b)/K.\nonumber
\end{align}
The remaining probability stays at $(o,b)$; terms are zero at their upper boundaries. Reward is realized before these transitions. For an additive expected-value objective, using expected current reward and the transition kernel is exact even though realized reward and learning can depend on the same case draw.

The counts belong to capability state $A$; overlap relative to this AI is derived from those capabilities. They are not independent extra relational coordinates. Problem state $P$ and the AI partner are held fixed in the baseline. Perfectly observed mastery and known transitions make the dynamic optimum an information benchmark. A practical policy may need a belief about mastery, as in Appendix~\ref{app:governance}.

\subsection{History experiments and analytical boundaries}
\label{app:histories}

The computational clock counts cases, not elapsed working time. The fixed-case-count experiment assigns one workflow for $n$ cases from the same initial state, propagates the complete state distribution, and evaluates both next workflows on that distribution. Current expected rewards are linear in $m_B$, so their comparison depends on its mean. Dynamic optimization, by contrast, uses the complete distribution where needed. With initial blind mastery $b_0$, fixed histories give
\begin{align}
\E[m_{B,n}^{AI}]&=K_B-(K_B-b_0)(1-\omega\eta/K)^n,\nonumber\\
\E[m_{B,n}^{IND}]&=K_B-(K_B-b_0)(1-\eta/K)^n.\label{eq:coverage_means}
\end{align}
These follow by taking expectations in the linear conditional transition. Expected current rankings reverse precisely when
\begin{equation}
\frac{\E[m_{B,n}^{AI}]}K<c<\frac{\E[m_{B,n}^{IND}]}K.
\label{eq:coverage_boundary}
\end{equation}
At baseline and $n=30$, the bounds are $.04000$ and $.1127225$. At the state level, independent work is strictly better when $m_B\geq2$; this occurs on 91.8034\% of independent-work histories and no AI-led histories. The historical contrast is mediated by mastery, not by an additional history label.

Table~\ref{tab:demo} reports the expected next-workflow values after the two thirty-case histories. These histories also produce different aggregate human accuracies: $62.18\%$ after AI-led work and $69.45\%$ after independent work. The equal-attainment experiment below separates composition from that difference in overall competence.

\begin{table}[htbp]
\centering\small
\begin{tabularx}{\textwidth}{@{}Yrrr@{}}
\toprule
\textbf{Prior workflow: 30 cases} &
\textbf{Human accuracy} &
\textbf{Next $I^{AI}$} &
\textbf{Next $I^{IND}$}\\
\midrule
\HistoryRows
\bottomrule
\end{tabularx}
\caption{Earlier workflows create different subsequent architectural fit. Values are model expectations from a common initial state and equal numbers of training cases. Each next workflow is evaluated over the full knowledge distribution created by the corresponding history. Next-case values include current effort but exclude the costs of producing the history; bold identifies the better next workflow.}
\label{tab:demo}
\end{table}

If $\omega=1$, the two learning kernels are identical, so fixed histories generate identical distributions and the interval in Eq.~\eqref{eq:coverage_boundary} is empty. Random mastery can still make state-responsive current choices useful; equal learning access removes \emph{architecture-created} distribution differences, not every possible value of observation. If $n=0$ or $\eta=0$, there is likewise no history difference. Very low or very high effort costs can make the same current workflow best after both histories.

The equal-attainment experiment stops each path when total mastery first reaches 17, without discarding paths. From $(10,1)$, AI-led learning at $\omega=0$ necessarily ends at $(16,1)$. Under independent work, the six new types are a uniform sample without replacement from ten unmastered AI-covered and four unmastered blind-spot types. Thus $m_B=1+Y$, where $Y$ is hypergeometric with population 14, four blind spots, and sample size six. Its expectation is $1+6(4/14)=19/7$. The resulting mean next independent-work value is $.8+(19/7)/25-.06=.8485714$.

Expected stopping times sum the expected waiting times for successive mastery gains: 42.2817 AI-led cases and 26.6853 independent-work cases. The latter incurs expected undiscounted effort cost $26.6853c=1.6011$. Equal final accuracy therefore isolates composition with unequal numbers of cases and unequal effort; it does not establish the superiority of either complete training policy. The separate fixed-case-count subgroup diagnostic remains in the outputs; it is not the equal-attainment experiment.

\begin{figure}[htbp]
\centering
\includegraphics[width=\textwidth]{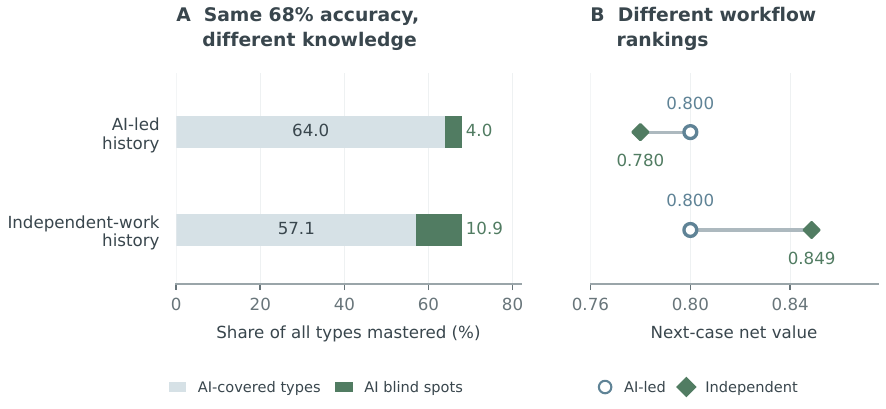}
\caption{\textbf{Learning histories can change architectural fit at equal human competence.} Every learning path reaches mastery of 17 of 25 types. (A) Earlier workflows create different knowledge profiles relative to the same 80\%-accurate AI. (B) Those profiles reverse the ranking of expected next-case net values. Fractional mastery counts are averages across paths; every path ends with exactly 17 mastered types. Training duration and effort differ across histories. Values are model expectations, not empirical estimates.}
\label{fig:demo}
\end{figure}

At the common $68\%$ human accuracy, independent work is worth $.78000$ after AI-led training and $.84857$ after independent training, while AI-led work remains worth $.80000$. Thus the preferred next workflow reverses at identical aggregate competence (Figure~\ref{fig:demo}).

\subsection{Policy classes, horizon, and terminal value}
\label{app:policy}

For a common initial state, discount $\gamma=.98$, and horizon $T$, we compare the best fixed workflow, the best schedule chosen before learning outcomes are observed, and the fully informed policy conditioned on time and mastery. A myopic state policy chooses the larger current net reward, breaking ties in favor of AI-led work. It is a separate comparator: it conditions on state but ignores investment returns. The optimized schedule is at least as good as the best fixed workflow, and the full optimum is at least as good as either; nesting guarantees only weak dominance, not a substantial strict gain.

Although the full process has 126 states, reward and the marginal transition of $m_B$ are independent of $m_O$. Hence control reduces exactly to six blind-mastery states for these workflows. Let $P_I^B$ denote their transition matrices and $R_I(b)$ their rewards. Backward induction is
\begin{equation}
V_t(b)=\max_{I\in\{I^{AI},I^{IND}\}}
\left\{R_I(b)+\gamma\sum_{b'}P_I^B(b,b')V_{t+1}(b')\right\},
\qquad V_T=U.
\label{eq:coverage_bellman}
\end{equation}
Full mastery observation is unnecessary: observing $m_B$ suffices for this optimum. Average unaided accuracy generally does not, since it combines $m_O$ and $m_B$.

\begin{table}[htbp]
\centering\small
\setlength{\tabcolsep}{4pt}
\begin{tabularx}{\textwidth}{@{}Yrrrr@{}}
\toprule
\textbf{Horizon / terminal rule} & \textbf{Myopic} & \textbf{Best schedule} & \textbf{Full} & \textbf{Full $-$ schedule}\\
\midrule
\PolicyRows
\bottomrule
\end{tabularx}
\caption{Expected discounted net value from $(10,1)$. In all six rows the best schedule is also the best fixed workflow: AI-led for $T=14$ with zero terminal reward, and independent work otherwise. ``Common continuation'' permits the same stationary myopic policy after $T$ for every class. Zeros denote numerical equality to within $10^{-12}$.}
\label{tab:policies}
\end{table}

The zero-terminal comparison in Table~\ref{tab:policies} charges all within-horizon effort and gives no value to mastery remaining at $T$. Over 30 cases, the full policy gains .351315 over myopic choice but only .007874 over the best schedule. The first difference largely reflects forward-looking learning investment. It should not be reported as the incremental value of observing state relative to an informed predetermined design.

To expose dependence on truncation, the second specification sets $U(b)$ to the infinite discounted value of the stationary myopic policy. With $a^{my}(b)$ its chosen workflow, $U$ solves
\begin{equation}
U(b)=R_{a^{my}(b)}(b)+\gamma\sum_{b'}P_{a^{my}(b)}^B(b,b')U(b').
\label{eq:coverage_terminal}
\end{equation}
This is a common feasible continuation, not a fitted terminal bonus or an infinite-horizon optimality claim. Every class may observe mastery at the boundary and follow that same policy thereafter; ``predetermined'' therefore refers to the first $T$ cases. Its terminal value is averaged over the full induced distribution. Independent work throughout the controlled interval attains the full optimum under all three reported horizons with this terminal rule. These results show that the feedback premium is sensitive to what happens after the study ends.

A further continuation experiment starts instead from each distribution produced by a 30-case history. With zero terminal reward and one or five cases remaining, optimal first actions differ: always AI-led after the AI-led history, versus independent work on 91.8\% of paths after independent work. At 14 or 30 cases remaining, independent work is the optimal first action at all states reached by either history. Different immediate fit can coexist with the same forward-looking investment choice. This is a horizon boundary on policy reversal, not a disappearance of the difference in knowledge.

\subsection{Sensitivity and alternative information conditions}
\label{app:sensitivity}

The initial parameter map crosses $c\in\{0,.02,\ldots,.16\}$, $\eta\in\{.25,.5,.75\}$, and $\omega\in\{0,.25,.5,.75,1\}$, holding all other primitives fixed. All 135 cells are supplied; 23 have a strict reversal of expected next-workflow rankings after 30-case histories. This count describes the declared grid, not a probability of reversal in real systems. Equation~\eqref{eq:coverage_boundary} supplies the underlying continuous boundary. Subsequent checks below were added after the baseline results; none replaces the baseline with a favorable parameter choice.

\paragraph{Granularity.}
Refining to $K=50,100,200$ scales initial mastery, AI coverage, and the history length proportionally, maintaining 80\% AI accuracy and equal exposure per type. After independent-work histories, the next independent-work values are .852455, .852322, and .852256, compared with .852723 at $K=25$. AI-led histories give .780000 in each case. The reversal survives refinement under this scaling. It is not a comparison at a fixed number of cases: the histories contain 60, 120, and 240 cases, respectively.

\paragraph{Imperfect combination.}
Let $\upsilon$ be the probability of retaining a correct human solution on an AI blind spot and $\phi$ the probability of losing an AI-correct answer through harmful override. Replacing the independent-work reward by
\[
R_{IND}(b)=.8(1-\phi)+\upsilon b/25-c
\]
while keeping learning fixed separates output integration from learning access. Six of nine checks with $\upsilon\in\{.6,.8,1\}$ and $\phi\in\{0,.02,.05\}$ retain the baseline fixed-case-count reversal. In other cells the loss of useful integration makes independent work unattractive after both histories. These checks do not model learning from disagreement or endogenous verification effort.

\paragraph{An equally accurate AI can change what human knowledge contributes.}
After equal-attainment training, replace $k$ of the AI's five blind-spot types with $k$ types drawn uniformly from its covered set, holding human mastery fixed. AI accuracy remains 80\%. The difference between the two histories' expected values under the \emph{same} independent-work continuation is
\[
.0685714\left(1-\frac{k}{4}\right),\qquad k=0,\ldots,5.
\]
It disappears at $k=4$ and changes sign at $k=5$. This tests the partner-specific nature of complementary knowledge; it does not show an optimal-workflow reversal or endogenize AI learning.

\paragraph{Selective review and the timing of information.}
Suppose a free flag, available before human effort, has sensitivity $s$ on AI blind spots and false-positive rate $f$ on AI-covered types, uniform within those classes. A third workflow uses independent work only when flagged. With ideal combination, its expected next-case value is
\begin{equation}
R_{SEL}(b)=.8+s\,b/25-c(.2s+.8f).
\label{eq:selective}
\end{equation}
Table~\ref{tab:selective} compares all three workflows after the equal-attainment histories. Weakly informative flags preserve the original choice contrast; a sufficiently informative flag makes selective review best after both histories. This is a one-case comparison under expanded information access, not a solution of a three-workflow learning problem.

\begin{table}[htbp]
\centering\small
\begin{tabularx}{\textwidth}{@{}rrrYrY@{}}
\toprule
 & & \multicolumn{2}{c}{\textbf{After AI-led history}} & \multicolumn{2}{c}{\textbf{After independent history}}\\
\cmidrule(lr){3-4}\cmidrule(l){5-6}
\textbf{$s$} & \textbf{$f$} & \textbf{$R_{SEL}$} & \textbf{Best choice} & \textbf{$R_{SEL}$} & \textbf{Best choice}\\
\midrule
\SelectiveRows
\bottomrule
\end{tabularx}
\caption{An informative, free pre-review flag can change the appropriate comparison. AI-led and unconditional independent-work values remain .800/.780 after AI-led training and .800/.84857 after independent training.}
\label{tab:selective}
\end{table}

A separate limiting calculation gives a perfect free failure flag ($s=1,f=0$) a fallback workflow with the independent-work learning kernel but expected effort cost $.2c=.012$. Covered-case learning still occurs through AI use; independent learning on blind spots occurs when the flag activates. This workflow dominates unconditional independent work. Over 30 cases with zero terminal value, always using fallback and the full optimum both earn 19.614906. The result identifies why information timing matters: the baseline pays for independence before knowing whether it is needed. It is not evidence that practical AI confidence estimates are free or perfectly informative.

\subsection{Exact solution and released materials}

Write the full learning kernels as $P_{AI}=I+L_O+\omega L_B$ and $P_{IND}=I+L_O+L_B$, where $L_O$ and $L_B$ act on different mastery coordinates. They commute, so the two kernels commute. The state distribution after any predetermined prefix depends only on elapsed time and the number of independent-work actions. This supports an exact dynamic program over $(t,d)$, with $d$ that count, to optimize all schedules without enumerating $2^T$ sequences. Although endpoint distributions depend only on counts, reward timing still matters and is retained in the recursion. A terminal value is integrated against the complete endpoint distribution, not evaluated at its mean.

The released code evaluates finite probability distributions exactly up to floating-point arithmetic, with no sampling or fitted parameters. Fourteen tests cover normalized transitions, the control reduction, Bellman/forward-evaluation agreement, analytical history expectations, attainment, refinement, partner reallocation, and selective review. The schedule solver matches the optimum from exhaustive enumeration of all $2^{14}=16{,}384$ schedules for three passive-learning rates under both terminal specifications.

The \texttt{computation/} directory supplies the model, experiments, tests, complete outputs, and the record of initial design choices and subsequent audits. The build script regenerates Figure~\ref{fig:demo} and Tables~\ref{tab:demo}, \ref{tab:policies}, and~\ref{tab:selective}. Reproducibility establishes the numerical illustration; differential learning access and its consequences require empirical tests.

\end{document}